\documentclass{article}
\usepackage[utf8]{inputenc}
\usepackage{fullpage}
\usepackage{graphicx}
\usepackage{subcaption}
\usepackage{helvet}
\usepackage[eulergreek]{sansmath}
\usepackage[table]{xcolor}
\usepackage{pgfplots}           
\usepackage{amsmath,amssymb}
\usepackage[hidelinks]{hyperref}
\usepackage[capitalize,noabbrev]{cleveref}
\usepackage{soul}
\usepackage{booktabs,tabularx,ragged2e}
\usepackage{xspace}
\usepackage[inline]{enumitem}
\usepackage{fnpct}
\usepackage{mhchem}
\usepackage{authblk}
\usepackage{siunitx}

\makeatletter
\newcommand*{\addFileDependency}[1]{
  \typeout{(#1)}
  \@addtofilelist{#1}
  \IfFileExists{#1}{}{\typeout{No file #1.}}
}
\makeatother

\usepackage{pgfplotstable}
\usepackage{float}
\usepackage[labelformat=empty, position=top]{subcaption}
\newcommand{\comment}[1]{}

\pgfplotsset{compat=1.14,
    /pgfplots/ybar legend/.style={
    /pgfplots/legend image code/.code={%
       \draw[##1,/tikz/.cd,yshift=-0.25em]
        (0cm,0cm) rectangle (3pt,0.8em);},
   },
}

\newcommand{\Rtwo}{R$^2$\xspace}
\newcommand{\COO}{CO$_2$\xspace}

\newcommand{\best}[1]{\textbf{#1}}

\makeatletter
\newcommand*{\etal}{%
    \@ifnextchar{.}%
        {et al}%
        {et al.\@\xspace}%
}
\makeatother

\title{Machine learning with persistent homology and chemical word embeddings improves prediction accuracy and interpretability in metal--organic frameworks} 
\author[1, 2]{Aditi S. Krishnapriyan\thanks{akrishnapriyan@lbl.gov}}
\author[2]{Joseph Montoya}
\author[3]{Maciej Haranczyk}
\author[2]{Jens Hummelshøj}
\author[1]{Dmitriy Morozov}
\affil[1]{Computational Research Division, Lawrence Berkeley National Laboratory, Berkeley, California 94720, United States of America}
\affil[2]{Toyota Research Institute, Los Altos, California 94022, United States of America}
\affil[3]{IMDEA Materials Institute, C/Eric Kandel 2, 28906 Getafe, Madrid, Spain}
\date{}

\begin{document}

\maketitle

\begin{abstract}
Machine learning has emerged
as a powerful approach in materials discovery.
Its major challenge is selecting features
that create interpretable representations of materials, useful across
multiple prediction tasks. We introduce an end-to-end machine learning model that automatically generates 
descriptors that capture a complex representation of a material's structure and chemistry. This approach builds
on computational topology techniques (namely, persistent homology) and word embeddings from natural
language processing. It automatically encapsulates geometric and chemical information directly from the material system.
We demonstrate our approach on multiple
nanoporous metal--organic framework datasets by predicting methane and carbon dioxide adsorption across different conditions. 
Our results show considerable
improvement in both accuracy and transferability across targets compared to
models constructed from the commonly--used, manually--curated features,
consistently achieving an average 25--30\% decrease in root-mean-squared-deviation
and an average increase of 40--50\% in R$^2$ scores.
A key advantage of our approach is interpretability:
Our model identifies the pores that correlate best to
adsorption at different pressures, which contributes to understanding atomic-level
structure--property relationships for materials design.
\end{abstract}

\section{Introduction}



Metal--organic frameworks (MOFs) exhibit properties beneficial for a
number of applications.  Their porosity and large internal surface areas make
them promising adsorbents for gas separation and storage;
their diverse chemistry leads to their use as catalysts
\cite{Rowsell2005, doi:10.1021/cr200190s, doi:10.1021/acscatal.8b04515}. The
number of MOF structures is massive --- there are thousands of experimentally
synthesized structures, but also many more hypothesized ones --- creating
a need for efficient tools and approaches to quickly identify MOFs best suited for a given applications.

The properties defining the best MOFs are dependent on the application.
For example, different gas adsorptions have different applications: for example, adsorption
of methane in the 65--5.8 bar range is
relevant to on-board vehicular natural gas storage technologies \cite{methanemofs2014}, while 
adsorption of carbon dioxide at lower pressure is important for \COO capture from flue gases \cite{doi:10.1021/cr2003272}. 

Molecular simulations have played an important role in the prediction of adsorption and diffusion behaviour of guest species in nanoporous materials. They have allowed computation of
Henry's coefficients, adsorption loadings and diffusion coefficients at various conditions \cite{odoh2015}.
But a larger challenge
remains: to advance our understanding of MOFs, it is necessary to recognize
geometric and chemical features responsible for their performance in particular
applications. These features offer useful clues for the design of new materials.

Machine learning offers a promising research direction to address this challenge.
ML techniques~\cite{jablonkareview2020, chongreview2020} have been used to
screen large databases of MOFs, and to predict their properties faster
than molecular simulations. Furthermore, feature representations developed for
ML help identify
correlations between MOF features and target properties. This makes
it possible to relate input features to a MOF's performance in a particular
application. To do so effectively, one needs to find interpretable feature
descriptors, whose values can be related to recognizable MOF properties \cite{doi:10.1021/acscombsci.5b00188, doi:10.1021/acscombsci.7b00056,
doi:10.1021/jacs.9b11084, moosavi2020, doi:10.1021/acs.chemmater.8b02257, shi2020machine}.
Additionally,
the diversity of properties and the vast number of structures makes it especially desirable to have an automatic
framework to generate expressive features that work across multiple
applications, enabling more transferability and less ``handcrafting.''

Creating a universal representation from the input material structure, suitable for all different
prediction tasks, is incredibly complicated. Typically, domain
experts select specific features as the model input, usually tailored to making predictions about a
particular property of interest. Often, this approach requires a large amount of manual processing to extract 
the necessary features \cite{simon2015best}.
%
For example, in the case of gas adsorption at high pressure,
guest molecules tend to occupy the entire void space in a material, so
void fraction can be used in predictive models. In contrast, for
gas adsorption at low pressures, the guest molecules aggregate in the strongly
binding regions of the material's pore---standard structural descriptors are not
able to capture this information as well.
Additionally, chemical interactions of the system, in particular local strong adsorption sites, are important in determining some gas
adsorption properties; this information also needs to be encoded in material
descriptors.


Besides geometry and topology, chemical makeup of the internal surfaces is
key for predicting MOF properties. Chemistry is
especially important for predicting adsorption capacities at low pressures.
Previous approaches have constructed chemical descriptors by incorporating
information from MOF building blocks, such as functional groups
\cite{doi:10.1021/acs.chemmater.8b02257, borboudakis_chemically_2017, anderson2020adsorption}. These approaches
have resulted in some improvements in predictive capabilities, but
they still require manual feature curation to inspect all of the building blocks in the dataset.
Moreover, the prediction accuracy of these descriptors often does not transfer
across structures and properties.


In this paper, we describe how to overcome the above challenges and present an end-to-end ML framework that 
automatically generates a material representation, while only requiring the basic
material structure (atomic coordinates and elemental composition) as input. As a consequence,
this approach avoids handcrafting representations that do not transfer
across property predictions. We use a topological descriptor,
called persistent homology~\cite{edelsbrunner2007},  to compute multi-scale signatures of the channels and voids in the pores of the material.
There have been previous approaches applying topological data analysis to materials \cite{lee2017, sorensen2020revealing}; however, in this work,
we show that descriptors can be constructed from topological data analysis for downstream machine learning tasks for materials.

Additionally, we use features built using word embedding techniques~\cite{wordembeddings2019} to describe chemical information.
As we demonstrate, 
this automated ML framework beats the standard structural descriptors in
predicting a variety of materials properties. We also show that the overall methodology --- coupling these features with ML
algorithms that assign importances --- opens the proverbial ML black box and
allows us to interpret the predictions by identifying geometric and chemical properties relevant to different
tasks.

\section{Methods}

\subsection{Datasets}
\label{sec:datasets}

We demonstrate our approaches on three datasets corresponding to MOFs of various diversity, and across a range of CH$_4$ and \COO uptake pressures predicted using grand cannonical Monte Carlo simulations \cite{wilmer_large-scale_2012,boyd_generalized_2016, coremofs2019}. 
The first dataset is the hypothetical MOFs (hMOFs) database generated by Wilmer
\etal \cite{wilmer_large-scale_2012}. The hMOF structures were taken from
MOFDB ({\url{http://hmofs.northwestern.edu}}),
which also has adsorption uptakes for carbon dioxide at five different pressures ranging from 0.05
bar to 2.5 bar.

The second dataset is the Boyd--Woo predicted MOF database  \cite{boyd_generalized_2016}
with the predicted methane and carbon dioxide adsorption capacities at low and high pressure, and methane and
carbon dioxide Henry's coefficients. The Henry's coefficients are expressed in terms of their logarithms.


Finally, we also included the 2019 CoREMOF dataset of the experimentally synthesized MOFs \cite{coremofs2019}.

For each structure in our dataset, as in our previous work
\cite{krishnapriyan2020}, we have determined the values of the following
commonly--used geometric descriptors. We call these structural descriptors, and
use them as a baseline to compare against topological descriptors:
\begin{enumerate*}[label=(\alph*)]
    \item
        pore limiting diameter (PLD), in (\AA), the diameter of the largest sphere to percolate through a material;
    \item
        largest cavity diameter (LCD), in (\AA), the diameter of the largest sphere than can fit inside the material's pore system;
    \item
        crystal density ($\rho$), in (kg/m\textsuperscript{3});
    \item
        accessible volume (AV), in (cm\textsuperscript{3}/g);
    \item
        accessible surface area (ASA), in (m\textsuperscript{2}/cm\textsuperscript{3}).
\end{enumerate*}
The values for these descriptors were computed using the Zeo++ software package \cite{zeoplusplus}.

\subsection{Automated topology--processing pipeline}
\label{sec:pers-hom}

\begin{figure}
    \centering
    \includegraphics[page=4]{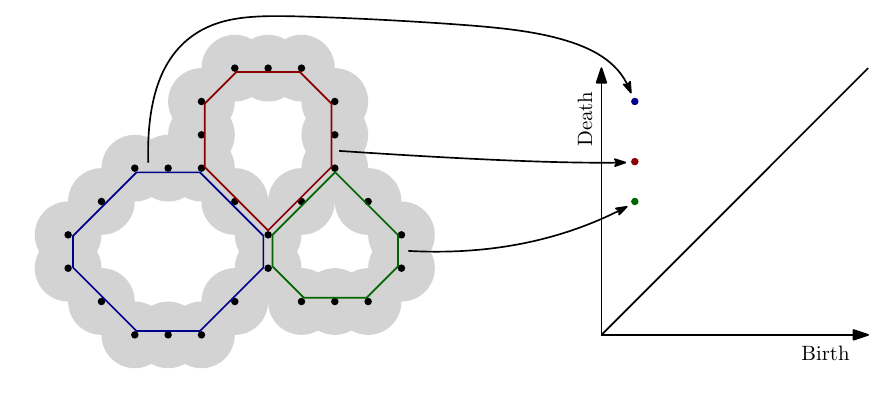}
    \caption{\textbf{Schematic outlining point cloud to persistence diagram.} (left) A point set (representing atomic centers) with balls of increasing
             radius around the points, (right) 1-dimensional persistence diagram of the
             point set. Representative cycles, corresponding to the points in
             the diagram, are highlighted with matching colors. The larger the loop, the higher the persistence value ($death-birth$). Figure created with Ipe 7.2.23 ({\url{http://ipe.otfried.org/}}).}
    \label{fig:persistence}
\end{figure}


We construct an automated pipeline to process an input MOF. We describe the topological structure of the MOFs using \emph{persistent homology} \cite{edelsbrunner2007}.
To normalize the size of each MOF, expressed as (periodic) base cells of
different sizes, we fill a (100\AA)$^3$ cell with the atoms of the MOF. The size
is chosen to be large enough to capture the statistics of the distribution of
the topological features in every structure.

We represent a MOF as a union of hard spheres centered on its atoms. We increase the
radii of these spheres and keep track of the changes in the topology of their
union. The changes come in two types: a topological feature, such as a loop or
a void, either appears or disappears. An important consequence of the
algebraic formulation of this process is that these events can be
paired uniquely, resulting in a set of birth--death pairs of radii, called a
\emph{persistence diagram}; see \cref{fig:persistence}. There are two persistence diagrams
relevant to us: a diagram that tracks births and deaths of loops that we
interpret as tunnels in the MOF (we call these 1-dimensional features), and a
diagram that tracks voids that we think of as pockets in the MOF (2-dimensional
features).  The difference in birth--death values is called
\emph{persistence} of the pair. Pairs of larger persistence capture more
prominent pores in the MOF. We compute persistence diagrams using the Dionysus
library ({\url{https://github.com/mrzv/dionysus}}).

Persistence diagrams are translated into vectors suitable as input for machine learning algorithms
via a modification of \emph{persistence images}, introduced by Adams \etal~\cite{adams_jmlr_2017}. 
The birth--death pairs $(b,d)$ are transformed
into birth--persistence pairs $(b, d-b)$. They are then convolved with Gaussians
and discretized onto a grid of a fixed size, by integrating the resulting
mixture of Gaussians in the cells of the grid. For this, 
 we use the resolution of $50 \times 50$ and a Gaussian spread of $\sigma = 0.15$. 


%


\subsection{Word embeddings}
\label{sec:we}

We incorporate word embeddings of the chemical elements to represent a given MOF's stoichiometric formula into our automated pipeline. 
We use this to capture the MOF's chemical information. The chosen embeddings were constructed from a
large corpus of abstracts with the word2vec algorithm \cite{wordembeddings2019}. The only input required is the elemental composition of the MOFs. 
Using word embeddings maintains the automated nature of our machine learning pipeline.
While the use of word embeddings to featurize composition do represent
an implicit knowledge that the chemical elements are distinct, they use no explicit element-specific properties 
and are themselves derived from an unsupervised learning procedure on raw text.
From an input MOF structure, we construct features based on the composition of each MOF structure that
represent word embeddings for the different elements in the MOF using the 
``matscholar\_el'' preset ElementProperty featurizer in
matminer \cite{ward_matminer_2018}. The features correspond to 200 embedding dimensions,
with the minimum, maximum, range, mean, and standard deviation for each dimension, for a total of 1000 values. 
We note that the different datasets have different numbers of unique elements. For example, the hMOF dataset has eight, while
the BW dataset has 16.


\subsection{Machine learning}

We use random forest \cite{breiman_rf_2001} regression to predict carbon dioxide and methane adsorption uptakes at different pressures including infinite dilution (the Henry's coefficients). One of our  motivations for using the random forest is the ability to determine the feature importances in the model. The random forest algorithm builds an ensemble of decision trees and chooses a random subset of features for each one. The frequency with which a particular feature is chosen for a split is an estimate for the importance of the said feature.

We build trees for different groups of features: topological features, standard structural features, word embeddings, a combined model of topological features and word embeddings, a combined model of topological and structural features, and a combined model of topological features, structural features, and word embeddings. The topological features consist of both the 1D and 2D persistence images. We train the random forest on the specific target prediction of each material. Each of the forests consists of 500 trees, and the final prediction is the average of the prediction of all trees in the forest. After training the random forest on a training set, predictions are made on an unseen test set. For most of the predictions, we use an 80\%/20\% training-test split. The quality of the prediction is evaluated by comparing the predicted adsorption values and the correct adsorption values. We quantify our predictions by computing the root-mean-square deviation, $\sqrt{\sum (\hat{y}_i - y_i)^2 / n}$, and the coefficient of determination (\Rtwo), $1 - \sum(y_i - \hat{y}_i)^2) / \sum (y_i - \bar{y})^2$.  
We also note that there are other approaches to utilize persistence diagrams in machine learning algorithms, 
such as by directly processing the diagrams through an input persistence layer in a neural network \cite{swenson2020persgnn}.

\subsection{Interpretability and representative cycles}
\label{sec:rep-cycles}

The algorithm used to compute persistence \cite{elz-tps-02} tracks
cycles that represent the topological features summarized in the persistence
diagram. The cycles are not unique, but they reveal the
atomic structures responsible for particular birth--death pairs.
In a crystal structure, representative cycles correspond to channels or voids in
the material.  We visualize the cycles to better understand the topological
features that appear in the MOFs.
We choose which cycle to
visualize using the feature importances found by the machine learning
algorithms. We compute the representative cycles using the aforementioned
Dionysus and visualize them with Zeo++ and VisIt.

\section{Results}

We evaluate the accuracy of the automatically generated descriptors for our machine learning models by predicting a number
of different targets across the different datasets. For each target, we
calculate the root-mean-square deviation (RMSD) and coefficient of determination
(\Rtwo score). For each target and each dataset, we include results from
models trained on only the topological features, only the word embeddings, and
both the topological features and the word embeddings (T + WE). We also include results
from the structural descriptors, described in \cref{sec:datasets},
as a baseline. Finally, we incorporate the standard structural descriptors by including models combining topological and structural descriptors (T + S), as well as topological descriptors, structural descriptors, and word embeddings (T + S + WE).


\subsection{hMOF dataset}


For the hMOF dataset, we predict carbon dioxide adsorption capacities at
different pressures, as shown in \cref{fig:hmof_CO$_2$_metrics}. The RMSD is low at
lower pressures because the distribution of carbon dioxide adsorption capacity
has low variance in this regime. While the topology-based model outperforms the
word embeddings, the model combining the two performs even better. We also see that the topological features always outperform the
structural features, often significantly. The word embeddings do not perform as well here. This is
likely due to the hMOF dataset lacking compositional diversity: the hMOF data
set contains only eight unique elements. 
Nevertheless, word embeddings help boost
the overall model performance when combined with the topological features.

We achieve the best performance by combining all three features together, but the accuracy achieved by subsets of the features is revealing.
Adding structural to topological features slightly improves the performance,
but doesn't match that of all three features combined.
On the other hand, the T + WE model performs
only slightly worse than the T + S + WE model, indicating that the topological
features capture most of the information that the structural features provide.


\pgfplotsset{
    cycle list={
    {blue, mark=*, very thick}, 
    {brown, mark=*,   very thick},
    {magenta, mark=*, very thick},
    {orange,  mark=*, very thick}, 
    {cyan,  mark=*, very thick},  
    {black!60!green, mark=*, very thick},
    }
}

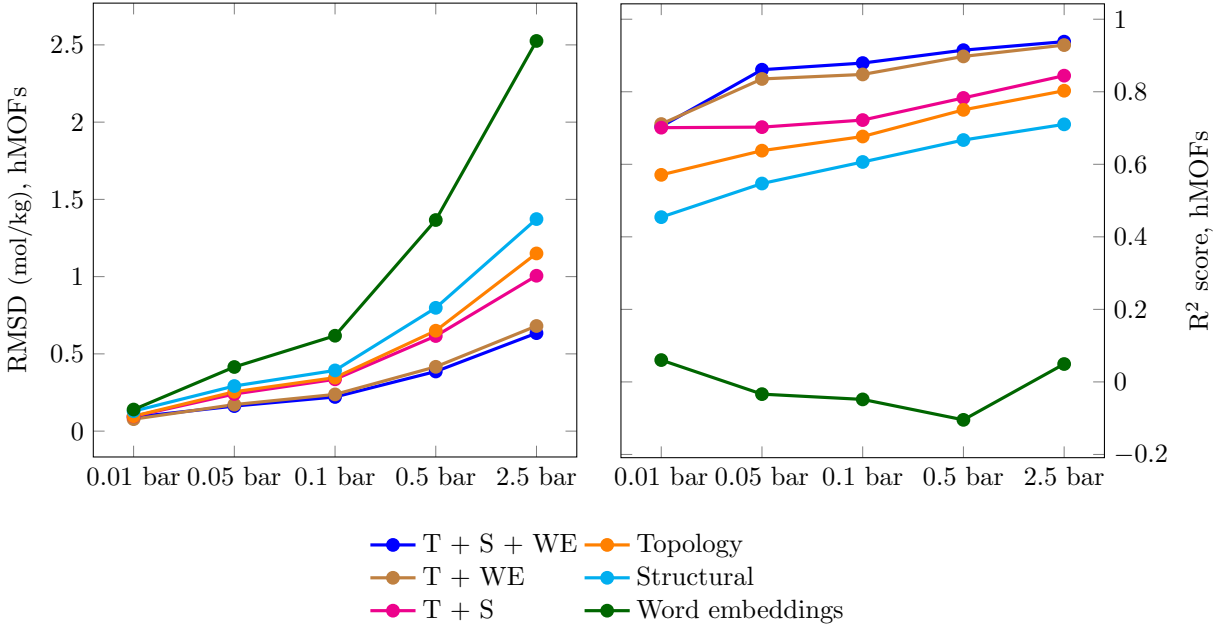
\begin{figure}[H]
\begin{minipage}{0.49\textwidth}
\centering
\begin{tikzpicture} 
\begin{axis}[
            symbolic x coords={0.01, 0.05, 0.1, 0.5, 2.5},
            xtick={0.01, 0.05, 0.1, 0.5, 2.5},
            xticklabels={0.01 bar, 0.05 bar, 0.1 bar, 0.5 bar, 2.5 bar},
                height=3in,
                width=.99\textwidth,
                every axis plot/.append style={thick},
                ylabel={RMSD {\small (mol/kg)}, hMOFs},
                legend columns=2,
                legend to name=rmsd-legend,
                legend style={draw=none},
                ]
        \addplot    +[] table[x=target, y=geometric-topology-we]
        {feature_comparisons/rmsd_hmofs_co2_mofs_v2.txt}; \addlegendentry{geometric-topology-we}
        \addplot    +[] table[x=target, y=combo]
        {feature_comparisons/rmsd_hmofs_co2_mofs_v2.txt}; \addlegendentry{combo}
        \addplot    +[] table[x=target, y=geometric-topology]
        {feature_comparisons/rmsd_hmofs_co2_mofs_v2.txt}; \addlegendentry{geometric-topology}
        \addplot    +[] table[x=target, y=topology]
        {feature_comparisons/rmsd_hmofs_co2_mofs_v2.txt}; \addlegendentry{topology}
        \addplot    +[] table[x=target, y=geometric]
        {feature_comparisons/rmsd_hmofs_co2_mofs_v2.txt}; \addlegendentry{geometric}
        \addplot    +[] table[x=target, y=we]
        {feature_comparisons/rmsd_hmofs_co2_mofs_v2.txt}; \addlegendentry{we}
\end{axis}
\end{tikzpicture}
\end{minipage}
\begin{minipage}{0.49\textwidth}
\centering
\begin{tikzpicture} 
\begin{axis}[
            symbolic x coords={0.01, 0.05, 0.1, 0.5, 2.5},
            xtick={0.01, 0.05, 0.1, 0.5, 2.5},
            xticklabels={0.01 bar, 0.05 bar, 0.1 bar, 0.5 bar, 2.5 bar},
                height=3in,
                width=.99\textwidth,
                every axis plot/.append style={thick},
                ylabel={\Rtwo score, hMOFs},
                ylabel near ticks,
                yticklabel pos=right,
                legend columns=3, transpose legend,
                legend to name=r2-legend,
                legend style={draw=none},
                legend cell align={left},
                ]
        \addplot    +[] table[x=target, y=geometric-topology-we]
        {feature_comparisons/r2_hmofs_co2_mofs_v2.txt}; \addlegendentry{T + S + WE} 
        \addplot    +[] table[x=target, y=combo]
        {feature_comparisons/r2_hmofs_co2_mofs_v2.txt}; \addlegendentry{T + WE} 
        \addplot    +[] table[x=target, y=geometric-topology]
        {feature_comparisons/r2_hmofs_co2_mofs_v2.txt}; \addlegendentry{T + S} 
        \addplot    +[] table[x=target, y=topology]
        {feature_comparisons/r2_hmofs_co2_mofs_v2.txt}; \addlegendentry{Topology} 
        \addplot    +[] table[x=target, y=geometric]
        {feature_comparisons/r2_hmofs_co2_mofs_v2.txt}; \addlegendentry{Structural} 
        \addplot    +[] table[x=target, y=we]
        {feature_comparisons/r2_hmofs_co2_mofs_v2.txt}; \addlegendentry{Word embeddings} 
\end{axis}
\end{tikzpicture}
\end{minipage}

\begin{center}
\ref{r2-legend}
\end{center}
\vspace*{-3mm}
\caption{\textbf{Model performances for hMOF dataset and \COO adsorption.} Comparison of root-mean-square deviation (left),
         coefficient of determination (right) in predicting gas uptakes in
         CO$_{2}$ for different features at different pressures for the hMOF
         dataset. The RMSD is low at
lower pressures because the distribution of carbon dioxide adsorption capacity
has low variance in this regime. The topological features consistently outperform the standard
         structural features at all pressures. The T + WE and T + S + WE models achieve the best performance in general.
         }
\label{fig:hmof_CO$_2$_metrics}
\end{figure}

\begin{table}[h]
	\centering
	\begin{tabular}{|c | c | c | c | c | c |}
	\rowcolor{lightgray}
	\hline
	Descriptor & 0.01 bar & 0.05 bar & 0.1 bar & 0.5 bar & 2.5 bar \\
	Structural & 0.45 & 0.55 & 0.61 & 0.67 & 0.71\\
	Topological & 0.57 & 0.64 & 0.68 & 0.75 & 0.80 \\
	T + S & 0.70 & 0.70 & 0.72 & 0.78 & 0.84 \\
	T + WE & \cellcolor{blue!25} 0.71 & 0.84 & 0.85 & 0.90 & 0.93 \\
	T + S + WE & 0.70 & \cellcolor{blue!25} 0.86 & \cellcolor{blue!25} 0.88 & \cellcolor{blue!25} 0.92 & \cellcolor{blue!25} 0.94 \\
	Best model, Fanourgakis \etal \cite{doi:10.1021/jacs.9b11084} & -- & 0.65 & -- & 0.90 & 0.93  \\
	\hline
	\end{tabular}
	\caption{\textbf{Summary of model performances for hMOF dataset and \COO adsorption.} Machine learning results for carbon dioxide adsorption predictions on the hMOF dataset at different pressures, represented by \Rtwo score. The best performing model for a given pressure is highlighted.}
	\label{tab:hmofs-predictions}
\end{table}


We compare our results to Fanourgakis \etal~\cite{doi:10.1021/jacs.9b11084}, who used standard structural features and a customized 
featurization based on atom types to predict CO$_{2}$ adsorption capacity
in the hMOF dataset. \cref{tab:hmofs-predictions} shows results for each of our models at different pressures,
along with the best model from Fanourgakis \etal~\cite{doi:10.1021/jacs.9b11084}.

Our model does particularly well at low pressures, achieving an $R^{2}$ score
of 0.86 at 0.05 bar, compared to 0.65 from \cite{doi:10.1021/jacs.9b11084}.
Carbon dioxide adsorption at low pressure has an important application:
carbon capture from flue gases. Thus, it is particularly promising to have a
generalized framework for accurate prediction of these targets. In general,
our model transfers well across different pressures, as demonstrated by
consistently high performance.



\subsection{BW dataset}






We evaluate the accuracy of the automated machine learning pipeline on the BW dataset.
We predict six targets grouped into three categories: the Henry's coefficient (log(K$_{H}$)) for CO$_2$ and CH$_4$, the gas uptakes for CO$_2$
at 0.15 and 16 bar, and the gas uptakes for CH$_4$ at 5.8 and 65 bar. 

\begin{table}[h]
    \centering
    \begin{tabular}{| l || r | r | r | r || r | r | r | r |}
    \hline
        & \multicolumn{4}{c||}{RMSD} & \multicolumn{4}{c|}{\Rtwo score}  \\
        \cline{1-9}
        Target        & S & T & T + WE & $\Delta$   & S & T & T + WE & $\Delta$ \\
    \hline
    log(K$_{H}$) \COO    & 0.46 & 0.38         & \best{0.33} & 28.3\%     & 0.60 & 0.68            & \best{0.78}  & 30\%     \\
    log(K$_{H}$) CH$_4$   & 0.27 & 0.20        & \best{0.18} & 33.3\%      & 0.50 & 0.73            & \best{0.79}  & 58\%    \\
    0.15 bar \COO             & 0.71 & 0.56          & \best{0.49} & 31\%      & 0.57 & 0.71            & \best{0.79}  & 38.6\%    \\
    16 bar \COO           & 1.9 & 2.53   & \best{1.80} & 5.3\%     & 0.93 & 0.88     & \best{0.94}  & 1.1\%    \\
    5.8 bar CH$_4$            & 19.18 & 14.85  & \best{13.97} & 27.2\%      & 0.68 & 0.82            & \best{0.84}  & 23.5\%    \\
    65 bar CH$_4$              & 23.87 & 20.61   & \best{17.66} & 26\%      & 0.83 & 0.87     & \best{0.90}         & 8.4\%      \\ 
    \hline
    \end{tabular}
    \caption{\textbf{Model performance on BW dataset.} Root-mean-square-deviation (RMSD) and 
         coefficient of determination (\Rtwo score) results in predicting the Henry's coefficient (log k$_{H}$) for CO$_{2}$ and CH$_{4}$, gas uptakes for CO$_{2}$, and gas uptakes for CH$_{4}$ for the BW dataset. Different sets of features (S = baseline structural, T = topological, T + WE = topological and word embeddings) are shown. For each target, the units are mol kg$^{-1}$ Pa$^{-1}$ and V$_{STP}$/V respectively. The best model is in bold. As the improvement from the topology + word embeddings is always greater than the structural features, the percentage of improvement (decrease in the case of RMSD and increase in the case of \Rtwo score) is also shown ($\Delta$).}
    \label{tab:bw-predictions}
\end{table}

\cref{tab:bw-predictions} shows the results of these predictions for the BW dataset for the baseline structural features (S), topological features (T), and topological features + word embeddings (T + WE). 
As a general trend, the T + WE model outperforms the structural features by a large amount, with an average (across all targets) decrease of 25.2\% in RMSD and an average increase of 26.6\% in \Rtwo score.
This is especially apparent for the Henry's coefficient predictions and the CO$_2$ and CH$_4$ gas uptakes at low pressure.
For these low pressure and infinite dilution gas adsorption predictions, to our knowledge, these topological descriptors are currently the best-performing
descriptors that only take into account geometric information about the MOF.
Supplementary \cref{fig:bw20k_metrics} shows further visualization of the results with different sets of features.

\subsection{CoREMOF dataset} 


Finally, we evaluate the accuracy of the automated ML pipeline on the CoREMOF dataset. To narrow the dataset in a principled manner, we only include MOFs with a known topology net \cite{li2014topological}, with each topology net appearing at least 15 times in the dataset for a total of approximately 50 topology nets in the whole dataset. We predict four targets here: the Henry's coefficient (log(K$_{H}$)) for CO$_2$ and CH$_4$ and the gas uptakes for CH$_4$ at 5.8 and 65 bar. 

\begin{table}[h]
    \centering
    \begin{tabular}{| l || r | r | r | r || r | r | r | r |}
    \hline
        & \multicolumn{4}{c||}{RMSD} & \multicolumn{4}{c|}{\Rtwo score}  \\
        \cline{1-9}
        Target        & S & T & T + WE & $\Delta$   & S & T & T + WE & $\Delta$ \\
    \hline
    log(K$_{H}$) \COO    & 0.90   & 0.73        &  \best{0.60}  & 33.3\%     &  0.26 &      0.53       & \best{0.69}  & 165\%   \\
    log(K$_{H}$) CH$_4$   & 0.34  &    0.30     &  \best{0.24} &   29.4\%   &  0.55 &      0.65      & \best{0.78}  &  41.2\%   \\
    5.8 bar CH$_4$            & 27.15 &  22.00  &  \best{20.19} &  25.7\%     & 0.47 &     0.65       & \best{0.71}  & 51.1\%    \\
    65 bar CH$_4$              & 32.06 &  25.57  &  \best{24.57} &   23.1\%  & 0.76 &   0.85   & \best{0.87}         & 14.5\%     \\ 
    \hline
    \end{tabular}
    \caption{\textbf{Model performance on CoREMOF dataset.} Root-mean-square-deviation (RMSD) and 
         coefficient of determination (\Rtwo score) results in predicting the Henry's coefficient (log k$_{H}$) for CO$_{2}$ and CH$_{4}$ and gas uptakes for CH$_{4}$ for the CoREMOF dataset. Different sets of features (S = baseline structural, T = topological, T + WE = topological and word embeddings) are shown. For each target, the units are mol kg$^{-1}$ Pa$^{-1}$ and V$_{STP}$/V respectively. The best model is in bold. As the improvement from the topology + word embeddings is always greater than the structural features, the percentage of improvement (decrease in the case of RMSD and increase in the case of \Rtwo score) is also shown ($\Delta$).}
    \label{tab:coremof-predictions}
\end{table}

The improvement in using our ML framework in contrast to the commonly used structural features is particularly apparent in prediction improvement for the Henry's coefficient's of both \COO and CH$_4$ as well as low pressure CH$_4$. This improvement is especially noticeable in \Rtwo scores. For example, as seen in \cref{tab:coremof-predictions}, our ML framework results in a 165\% improvement over the structural features when predicting the Henry's coefficient for \COO. The implications here are vast as adsorption in the infinite dilution regime, such as is commonly seen at low partial pressures, is very important for carbon capture applications. Moreover, the same model provides additional improvement over RMSD and \Rtwo scores across all the targets, with an average decrease of 27.8\% in RMSD and an average increase of 68\% in \Rtwo score. While the same structural features cannot be used for accurate predictions across many different targets, in contrast, our model shows far greater transferability. Supplementary \cref{fig:coremofid_metrics} shows further visualization of the results with different sets of features. 

Notably, across all the datasets, the model combining topological and structural features only performs marginally better than the topological features alone. This indicates that the topological features are capturing almost everything the structural features capture, as well as much more.

\section{Interpretability}

We also show the utility of our approach from an interpretability point of view. The feature importances extracted from the ML models contain important information to enhance our understanding of the material design process, and we explore multiple facets of this in the next sections.

\subsection{Feature analysis}




The random forest algorithm infers the importance of individual features by
measuring how frequently they are used by the decision trees to make a
prediction about a MOF. In our methodology, there are three distinct types of
features: topological, structural, and word embeddings.
Further, topological
features come in two types, 1-dimensional features that capture the distribution
of channels in the MOF and 2-dimensional features that describe the voids.
Each of those consists of 2500 individual features (pixels in the
persistence image), but we combine them to infer the aggregate importance of the
different feature types.
In this section, we analyze contributions from the topological and word
embedding features, since the structural features contribute little extra
information.

\begin{figure}[H]
\begin{minipage}{0.33\textwidth}
\centering
\begin{tikzpicture}
\begin{axis}[
                title={BW},
                ybar stacked,
                ymin=0, ymax=1,
                symbolic x coords={logKH_CO2, logKH_CH4, pure_uptake_CO2_298.00_15000, pure_uptake_CO2_298.00_1600000, CH4LPSTP, CH4HPSTP},
                xtick={logKH_CO2, logKH_CH4, pure_uptake_CO2_298.00_15000, pure_uptake_CO2_298.00_1600000, CH4LPSTP, CH4HPSTP},
                xticklabels={log(K$_{H}$) CO$_2$, log(K$_{H}$) CH$_4$, 0.15 bar CO$_2$, 16 bar CO$_2$, 5.8 bar CH$_4$, 65 bar CH$_4$},
                xticklabel style={rotate=75, anchor=east, align=right},
                height=2in,
                width=.99\textwidth,
            ]
    \addplot    +[black,fill=magenta!60!white] table[x=target, y=1d]                   {feature_summations/combo_bw20k_feature_imp_summations_6targets_restricted.txt}; 
    \addplot    +[black,fill=blue!60!white]    table[x=target, y=2d]                {feature_summations/combo_bw20k_feature_imp_summations_6targets_restricted.txt}; 
    \addplot    +[black,fill=red!60!white]     table[x=target, y=we]                {feature_summations/combo_bw20k_feature_imp_summations_6targets_restricted.txt}; 
\end{axis}
\end{tikzpicture}
\end{minipage}
\begin{minipage}{0.33\textwidth}
\centering
\begin{tikzpicture}
\begin{axis}[
                title={CoREMOFs}, 
                ybar stacked,
                ymin=0, ymax=1,
                symbolic x coords={logKH_CO2, logKH_CH4, CH4LPSTP, CH4HPSTP},
                xtick={logKH_CO2, logKH_CH4, CH4LPSTP, CH4HPSTP},
                xticklabels={log(K$_{H}$) CO$_2$, log(K$_{H}$) CH$_4$, 5.8 bar CH$_4$, 65 bar CH$_4$},
                xticklabel style={rotate=75, anchor=east, align=right},
                height=2in,
                width=.99\textwidth,
                legend style={draw=none},
                legend columns=1,
                legend style={at={(1.55,.6)}},
                legend cell align={left}
            ]
    \addplot    +[black,fill=magenta!60!white] table[x=target, y=1d]                   {feature_summations/combo_coremofid_feature_imp_summations_4targets_restricted.txt}; 
    \addplot    +[black,fill=blue!60!white]    table[x=target, y=2d]                {feature_summations/combo_coremofid_feature_imp_summations_4targets_restricted.txt}; 
    \addplot    +[black,fill=red!60!white]     table[x=target, y=we]                {feature_summations/combo_coremofid_feature_imp_summations_4targets_restricted.txt}; 
\end{axis}
\end{tikzpicture}
\end{minipage}
\begin{minipage}{0.33\textwidth}
\centering
\begin{tikzpicture}
\begin{axis}[
                title={hMOFs}, 
                ybar stacked,
                ymin=0, ymax=1,
                symbolic x coords={0.01, 0.05, 0.1, 0.5, 2.5},
                xtick={0.01, 0.05, 0.1, 0.5, 2.5},
                xticklabels={0.01 bar CO$_2$, 0.05 bar CO$_2$, 0.1 bar CO$_2$, 0.5 bar CO$_2$, 2.5 bar CO$_2$},
                xticklabel style={rotate=75, anchor=east, align=right},
                height=2in,
                width=.99\textwidth,
                reverse legend,
                legend style={draw=none},
                legend columns=3,
                legend to name=feature-sum,
                legend style={at={(1.55,.6)}},
                legend cell align={left}
            ]
    \addplot    +[black,fill=magenta!60!white] table[x=target, y=1d]                   {feature_summations/co2hmofs_combo_feature_imp_summations_restricted_0.2.txt}; \addlegendentry{1D topology}
    \addplot    +[black,fill=blue!60!white]    table[x=target, y=2d]                {feature_summations/co2hmofs_combo_feature_imp_summations_restricted_0.2.txt}; \addlegendentry{2D topology}
    \addplot    +[black,fill=red!60!white]     table[x=target, y=we]                {feature_summations/co2hmofs_combo_feature_imp_summations_restricted_0.2.txt}; \addlegendentry{Word embeddings}
\end{axis}
\end{tikzpicture}
\end{minipage}
\vspace{-1em}
\begin{center}
\ref{feature-sum}
\end{center}
\vspace{-1.5em}
\caption{\textbf{Feature analysis of machine learning models.} Summary of relative feature importance across different targets for
         the 1D, 2D topological features, and word embeddings. The BW, CoREMOF, and hMOF datasets are shown here.}
\label{fig:feature-importance-summary}
\end{figure}
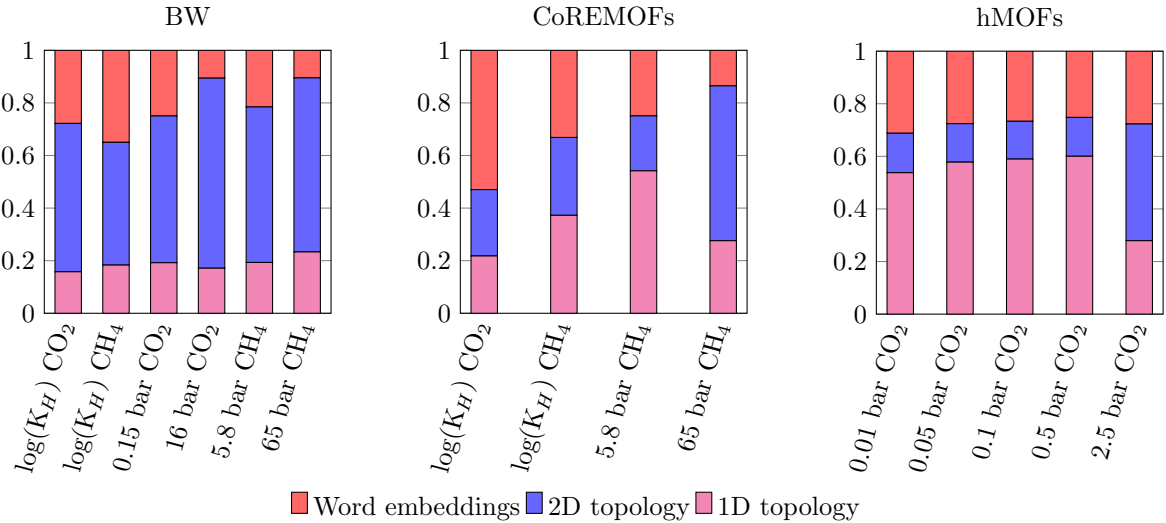

\cref{fig:feature-importance-summary} shows the relative importance of
topological descriptors and word embeddings.
For the BW dataset, 2D features are most important for the prediction, with
word embeddings playing a larger role in the predictions of the Henry's
coefficient. For the CoREMOF dataset, word embeddings are more important,
especially for the \COO Henry's coefficient
where they account for 50-60\% of the decisions, with topological features
dominating the importance of predictions for both low and high pressure methane adsorption (albeit, 1D features play a larger role in low pressure methane adsorption,
while 2D features play a larger role in high pressure methane adsorption).
For the hMOF dataset, 1D topological features are most important at low
pressures, with 2D being more important at higher pressure, and word embeddings
used in $\sim30\%$ of the decisions.

As \cref{fig:feature-importance-summary} shows, topological features play
a major role in predicting gas adsorption, with the 1-dimensional channels being
especially important for adsorption at low pressures in the CoREMOF and hMOF
datasets, and 2-dimensions voids being important for the predictions with the
BW dataset.
The differences in feature importances
can also be linked back to the data: for example, the
CoREMOF MOFs tend to have smaller pores than the BW MOFs. 


These results reveal the importance of different properties for different tasks.
They support the claim that chemical information is more important for infinite
dilution and low-pressure \COO adsorption. In these conditions, the
specific interactions between the gas and the MOF framework, e.g. manifested as strong binding sites, play an
important role in adsorption capacity --- the word embeddings capture this
non-structural information.
On the other hand, methane adsorption at higher pressure is mostly described by
2D topology features, which can described voids at large, a trend that we also observed in zeolites \cite{krishnapriyan2020}.

Our results also suggest why the conventional structural descriptors perform
especially poorly when predicting \COO adsorption in hMOFs at low pressure or in the infinite dilution region.
The standard structural features describe the pore geometry by the largest
sphere to percolate through the materials and the largest sphere that can fit
inside its pore system. At low pressures and/or in the infinite dilution region, the standard structural features 
are not able to capture the nuance of the gas molecules aggregating closer to the binding regions of the porous framework.
In contrast, topological features record the widths of the
channels that criss-cross the MOF as well as the sizes of different cavities. 
They also distinguishing between the distribution of channels and voids, by separating
1D and 2D topological features, and record other finer information about their shape.




\vspace{-0.5em}






\subsection{Topological features and representative cycles}

Nanoporous materials, and especially MOFs, are known for how tunable they are:
experimentalists can synthesize materials with precisely sized pores.
Understanding how structure features influence a particular material property helps
guide this process. Our approach incorporating persistent homology is especially helpful in this task.

The points in a persistence diagram correspond to voids and channels of specific
sizes. A point $(b,d)$ in a 2-dimensional diagram is generated by a cavity that
can fit the largest sphere of radius $d$; the largest sphere that can access the cavity has radius $b$.
A point $(b,d)$ in a 1-dimensional diagram is produced by a channel in the
material, specifically, by its narrowest ``bottleneck.'' The death value, $d$, records
the radius of the largest sphere that can pass through this bottleneck. The
birth value, $b$, records how close the atoms of the bottleneck are to each other.

For each dataset and each target property, the
most important 1D and 2D birth-death points, as identified by the random forest
algorithm, are listed in Table~\ref{tab:pore-sizes}. 
We note a few patterns. In the case of methane adsorption in all three
regimes (infinite dilution, low pressure, and high pressure), the 2D birth and
death values are similar for both the BW and CoREMOF datasets --- in fact, almost
identical for the infinite dilution and high pressure cases. Specifically, birth
values are around 2.3--2.4 \AA\ for high pressure methane adsorption, and 3.4--3.8 \AA\
for low pressure and infinite dilution methane adsorption. Death
values are 3.2 \AA\ for high pressure methane adsorption, and 4--4.6 \AA\ for
low pressure and infinite dilution methane adsorption. The radius of a methane molecule is assumed to be 3.8 \AA.
These results suggest that pores somewhat larger than this radius
adsorb well at low pressures and partial pressures,
while at high pressures slightly smaller pores influence the overall adsorption capacity of the MOF.

Another pattern to note
in the hMOF dataset is that 1D death values get larger as pressure
increases, meaning the size of the largest sphere able to pass through the
channel increases. The radius of a \COO molecule is assumed to be 3.3 \AA. For high pressure targets,
the model picks out the channels that can accommodate the molecule of this size. The 1D birth/death values 
for lower pressures correspond to smaller pores, such as the porous surface, which is related to the binding regions of the material's pore.

\begin{table}[h]

\begin{subtable}{1\textwidth}
\centering
\begin{tabular}{|c | c | c | c  | c|}
	\rowcolor{lightgray}
	\hline
	Target property  & 1D birth & 1D death & 2D birth & 2D death  \\
	log(K$_{H}$) CO$_{2}$ & 1 & 4 & 3.3 &  4.1 \\
	log(K$_{H}$) CH$_{4}$ & 1.6 & 2 & 3.6 & 4.4    \\
	0.15 bar CO$_{2}$ & 3.5 & 3.6 & 3.4 & 4 \\
	16 bar CO$_{2}$ & 1.7 & 2 & 3.1 & 3.9  \\
	5.8 bar CH$_{4}$ & 1.4 &3 & 3.8 & 4.6  \\
	65 bar CH$_{4}$ & 3.6 & 4.3 & 2.3 & 3.2 \\
	\bottomrule
	\end{tabular}
	\vspace{0.5em}
	\caption{(a) BW dataset.}
\label{tab:top-bd-bw}
\end{subtable}

\bigskip
\begin{subtable}{1\textwidth}
\centering
\begin{tabular}{|c | c | c | c  | c|}
	\rowcolor{lightgray}
	\hline
	Target property  & 1D birth & 1D death & 2D birth & 2D death  \\
	log(K$_{H}$) CO$_{2}$ & 0.3 & 1.3 & 2.3 &  3.1 \\
	log(K$_{H}$) CH$_{4}$ & 0.3 & 1 & 3.6 & 4.4    \\
	5.8 bar CH$_{4}$ & 1 & 3.3 & 3.4 & 4  \\
	65 bar CH$_{4}$ & 3.9 & 4.8 & 2.4 & 3.2 \\
	\hline
	\end{tabular}
	\vspace{0.5em}
	\caption{(b) CoREMOF dataset.}
\end{subtable}
\label{tab:top-bd-core}

\bigskip
\begin{subtable}{1\textwidth}
\centering
\begin{tabular}{|c | c | c | c  | c|}
	\rowcolor{lightgray}
	\hline
	Target property  & 1D birth & 1D death & 2D birth & 2D death  \\
	0.01 bar CO$_{2}$ & 0.02 & 0.7 & 3.2 &  3.5 \\
	0.05 bar CO$_{2}$ & 1.1 & 1.6 & 1.6 & 2.1    \\
	0.1 bar CO$_{2}$ & 1.1 & 2.7 & 4.4 & 5.5 \\
	0.5 bar CO$_{2}$ & 1.3 & 3.5 & 4.7 & 5.8  \\
	2.5 bar CO$_{2}$ & 1 & 3.7 & 4 & 5.1  \\
	\hline
	\end{tabular}
	\vspace{0.5em}
	\caption{(c) hMOF dataset.}
\end{subtable}
\label{tab:top-bd-hmof}
\vspace{-1em}
\caption{Most important 1D/2D birth--death points for the different datasets (in Angstroms). These values correspond to the porous framework sizes most important for a given adsorption task.} \label{tab:pore-sizes}
\end{table}

We can dissect topological representations further and extract representative cycles for
each point. Although these cycles are not unique --- we are at the mercy of
certain choices persistent homology calculation makes --- they are helpful in
visualizing the cavities and channel bottlenecks represented by the points in
the persistence diagram.

Since we train our machine learning algorithm on vectorized persistence images, we have to
take an extra step to identify the points in a persistence diagrams with
relevant representative cycles. We illustrate our steps for this approach in Supplementary Figure~\ref{fig:rep-cycle-schematic}.


\begin{figure}[H]
  \begin{minipage}[t]{0.02\textwidth}
    \textbf{(a)}
  \end{minipage}
  \begin{minipage}[T]{0.4\textwidth}
	\centering
	\includegraphics[width=0.9\textwidth]{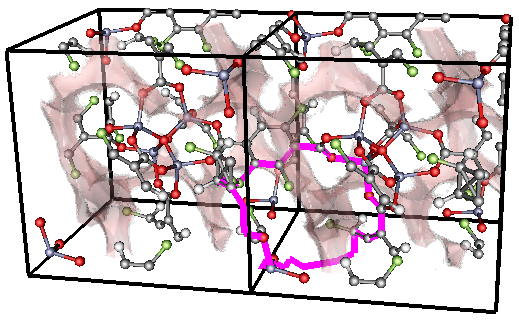}
\end{minipage}
  \begin{minipage}[t]{0.02\textwidth}
    \textbf{(b)}
  \end{minipage}
\begin{minipage}[T]{0.4\textwidth}
	\centering
	\includegraphics[width=0.9\textwidth]{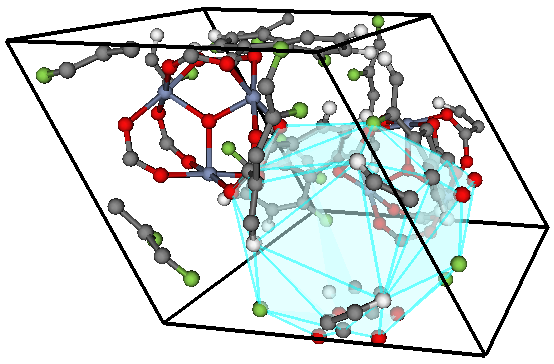}
\end{minipage}
\caption{\textbf{Example 1D and 2D representative cycles for different MOFs.} (a) 1D channel, hMOF-675 (hMOFs)  (b) 2D void, str-m4-o14-o14-acs-sym-5 (BW). The representative cycles
are picked based on the approach described in Supplementary Figure~\ref{fig:rep-cycle-schematic}. Figure created with VisIt 3.1.4 ({\url{https://wci.llnl.gov/simulation/computer-codes/visit}}).}
\label{fig:rep-cycles}
\end{figure}

\comment{
\begin{figure}[H]
  \begin{minipage}[t]{0.02\textwidth}
    \textbf{(a)}
  \end{minipage}
\begin{minipage}[T]{0.4\textwidth}
	\centering
	\includegraphics[width=0.9\textwidth]{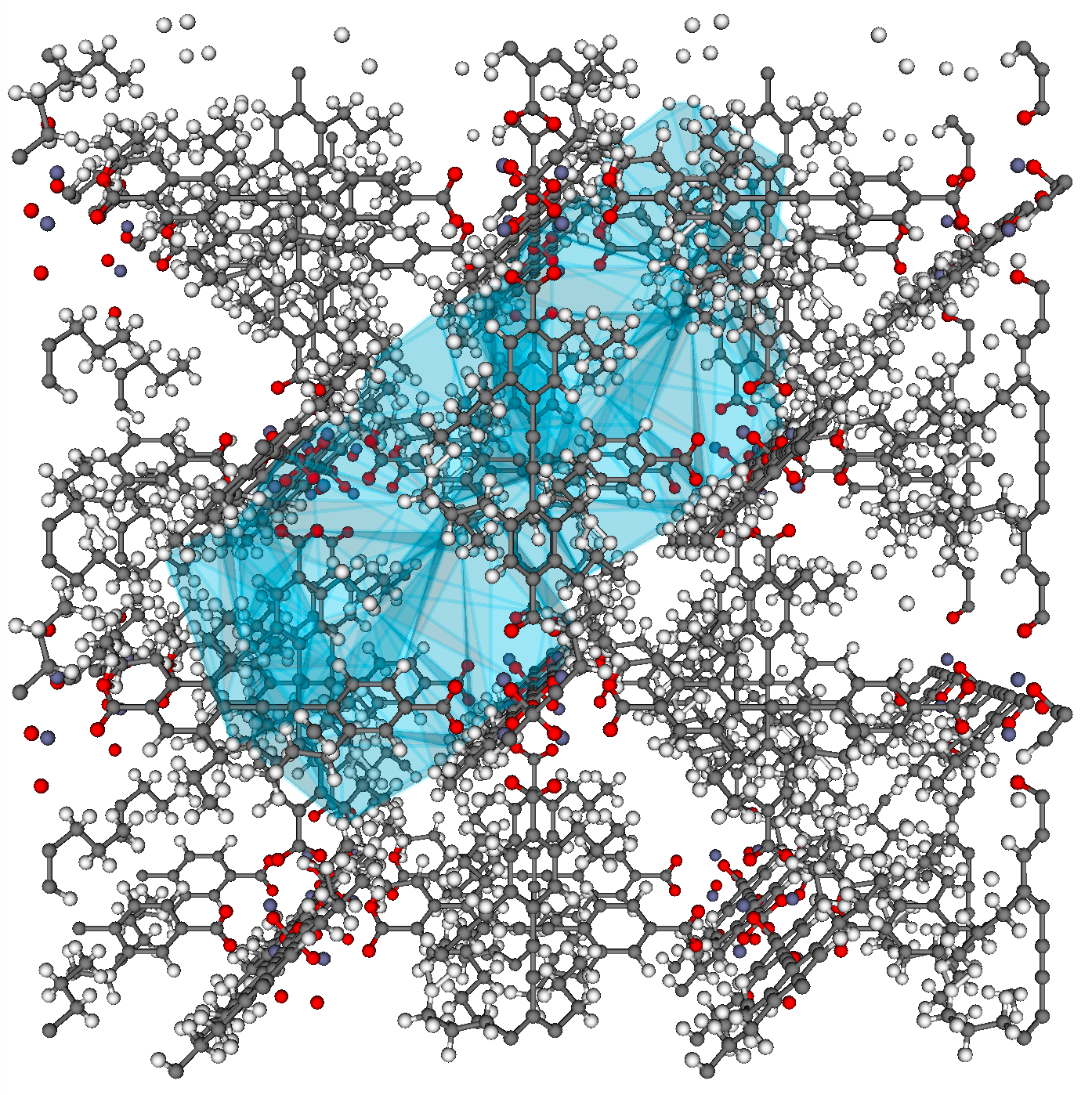}
\end{minipage}
  \begin{minipage}[t]{0.02\textwidth}
    \textbf{(b)}
  \end{minipage}
\begin{minipage}[T]{0.4\textwidth}
	\centering
	\includegraphics[width=0.9\textwidth]{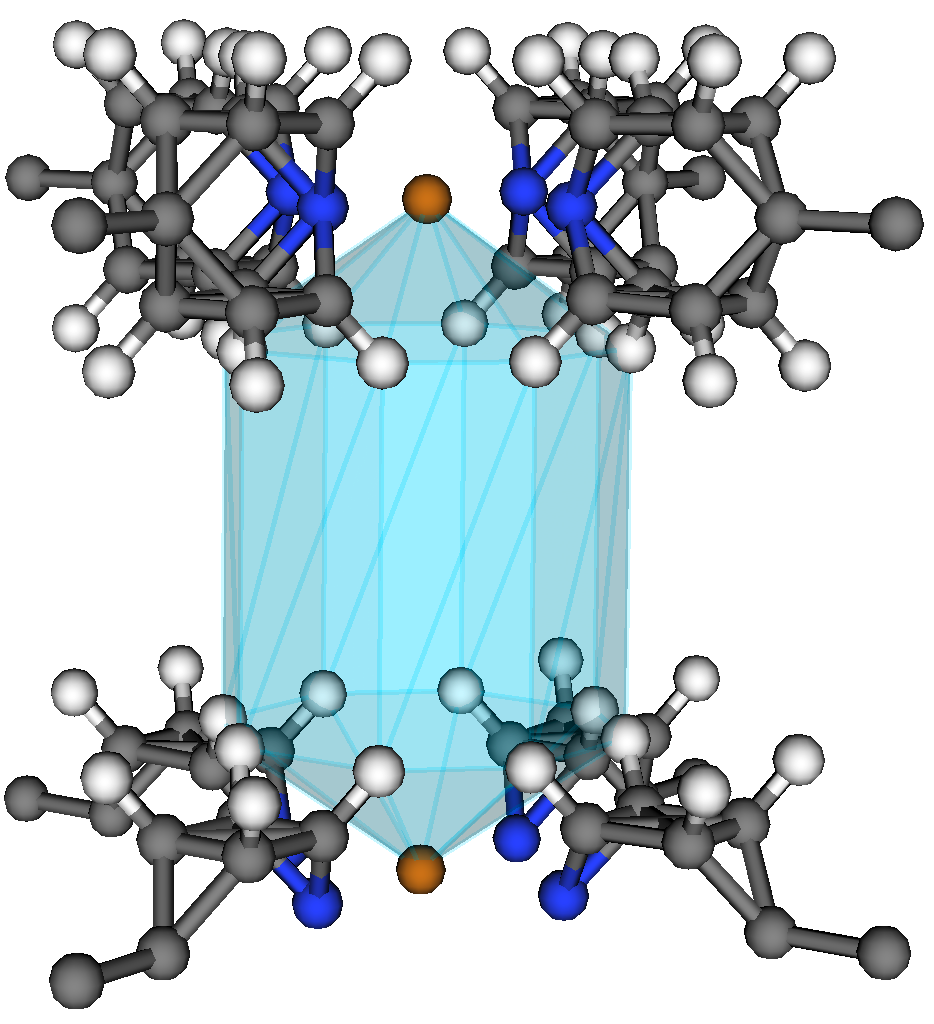}
\end{minipage}

\hfill

  \begin{minipage}[t]{0.02\textwidth}
    \textbf{(c)}
  \end{minipage}
\begin{minipage}[T]{0.4\textwidth}
	\centering
	\includegraphics[width=0.9\textwidth]{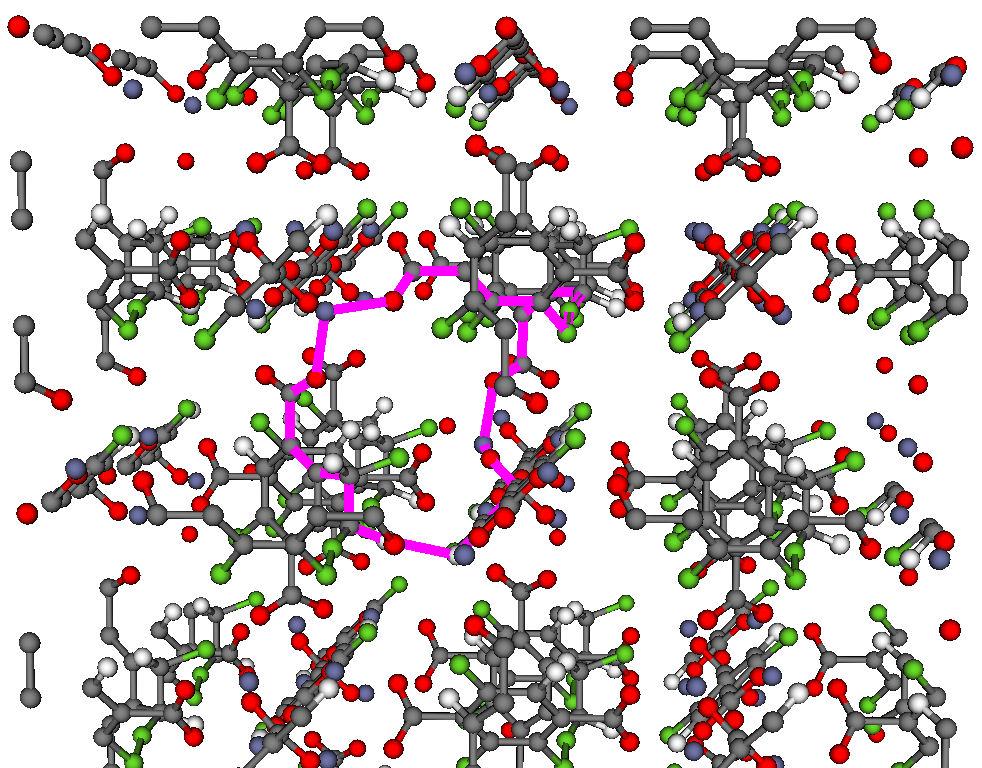}
\end{minipage}
  \begin{minipage}[t]{0.02\textwidth}
    \textbf{(d)}
  \end{minipage}
\begin{minipage}[T]{0.4\textwidth}
	\centering
	\includegraphics[width=0.9\textwidth]{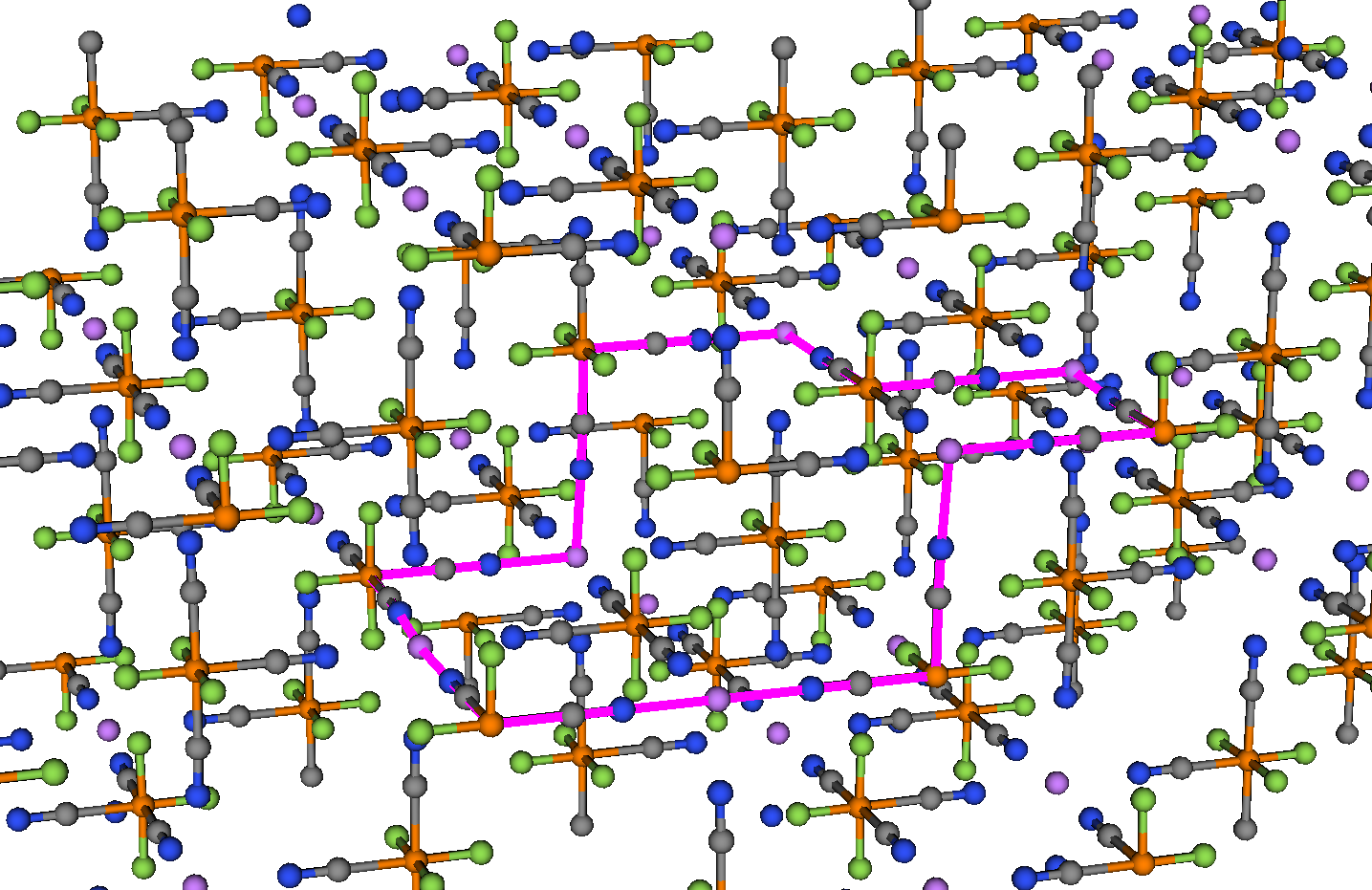}
\end{minipage}

\caption{\textbf{Representative cycles for different MOFs with high gas adsorption capacities at various pressures.} (a) 2D void, hMOF-3032 (hMOFs)  (b) 2D void, YEMTER (CoREMOFs) (c) 1D channel, hMOF-675 (hMOFs) (d) 1D channel, QABKIQ (CoREMOFs)  }
\label{fig:rep-cycles}
\end{figure}
}
We extracted the representative cycles from the high gas adsorption MOFs from
different databases. Two examples, including both 1D topology (channels) and 2D
topology (voids), appear in Figure~\ref{fig:rep-cycles}. One notable trend is that
the loop in Figure~\ref{fig:rep-cycles}(a) is present in many of the materials in the hMOF dataset that have
high CO$_{2}$ adsorption at low pressure. Similarly, the void size seen in Figure~\ref{fig:rep-cycles}(b) is present in many of the MOFs with high Henry's coefficients for \COO adsorption.

\begin{figure}[H]
\centering
  \begin{minipage}[t]{0.02\textwidth}
    \textbf{(a)}
  \end{minipage}
\begin{minipage}[T]{0.4\textwidth}
	\centering
	\includegraphics[width=0.9\textwidth]{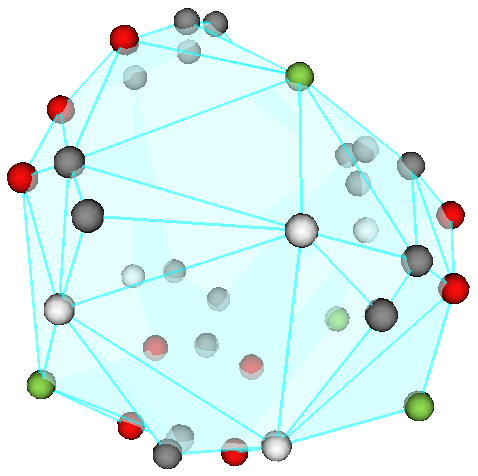}
\end{minipage}
  \begin{minipage}[t]{0.02\textwidth}
    \textbf{(b)}
  \end{minipage}
\begin{minipage}[T]{0.4\textwidth}
	\centering
	\includegraphics[width=0.9\textwidth]{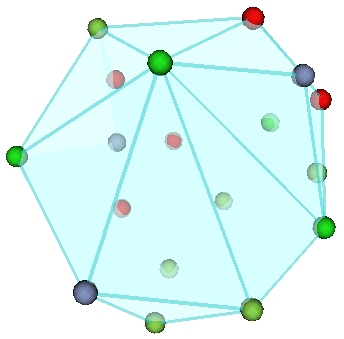}
\end{minipage}

\hfill

  \begin{minipage}[t]{0.02\textwidth}
    \textbf{(c)}
  \end{minipage}
\begin{minipage}[T]{0.4\textwidth}
	\centering
	\includegraphics[width=0.9\textwidth]{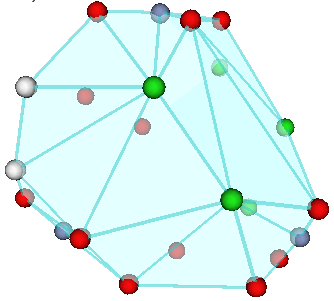}
\end{minipage}
  \begin{minipage}[t]{0.02\textwidth}
    \textbf{(d)}
  \end{minipage}
\begin{minipage}[T]{0.4\textwidth}
	\centering
	\includegraphics[width=0.9\textwidth]{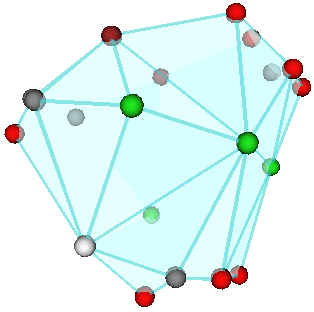}
\end{minipage}

\caption{\textbf{Correlating void structure to MOF property.} (a) str-m4-o14-acs-sym-8  (b) str-m4-o1-o22-acs-sym-94 (c) str-m4-o1-o24-acs-sym-96 (d) str-m4-o1-o24-acs-sym-165. The representative cycles of voids corresponding to the void most correlated with the \COO Henry's coefficient in example MOFs with high \COO Henry's coefficients. The voids are all composed of a similar bonding structure, with each different atom type represented by a different color. 
As noted in \cite{martin2012similarity}, the process of identifying the void structure that appears in top performing MOFs can be extremely time-consuming via manually detected features. Thus, we hope that our much faster and topologically--grounded approach will allow for further study in pinpointing the channel and void shapes and bonding structures that correlate best to important material's properties, thereby encouraging the targeted design of structures to maximize desirable properties. Figure created with VisIt 3.1.4 ({\url{https://wci.llnl.gov/simulation/computer-codes/visit}}). }
\label{fig:co2-cycles}
\end{figure}

We expand on the latter by showing, as an example, in Figure~\ref{fig:co2-cycles} the extracted voids that appear in a number of the top MOFs with a high \COO Henry's coefficient.
As noted in \cite{martin2012similarity}, the process of identifying the void structure that appears in top performing MOFs can be extremely time--consuming via manually detected features.
Thus, we hope that our approach will allow for further study in pinpointing the channel and void shapes and bonding structures that correlate best to important material's properties, thereby encouraging the targeted design of structures to maximize desirable properties. 

\subsection{Word embeddings and material properties}

We explore the interpretability of the word embeddings by relating their
importances in predicting MOF properties and in predicting chemical properties of
individual elements. The former we obtain from the random forests just
as the importances of the topological features. To calculate the importances
for individual elements, we retrieve word
embeddings for all the elements in the matscholar database
\cite{wordembeddings2019} and use these as features to train models to predict
various chemical properties --- electronegativity, atomic radius, electrical
resistivity, melting point, etc.\ --- of the pure elements contained in pymatgen's
`periodic\_table' module \cite{ONG2013314}. We extract the feature importances
for each of these models.
Because each MOF has 1000 features, summarizing the distribution of 200 features
over its elements, as described in \cref{sec:we}, we sum up
the MOF feature importances corresponding to the same elemental feature.

We take the subset of feature importances that account for 90\% of the random
forest decisions. By definition, these features describe the subspace of our input where
most of the decisions are made to make a prediction about the given target
property. Given a MOF target property and a chemical target property,
we compute the Jaccard similarity between the two subsets of features. This
metric measures the relative size of the subspace, important for the random
forest decisions for both targets.

\cref{tab:we-interpret} lists the top
three materials properties by similarity to each MOF target property; all of
them have a Jaccard similarity greater than 0.4. Following this procedure, we identify
the chemical property with the strongest informational relevance to
a given MOF target property.
\begin{table}[h]
	\centering
	\begin{tabular}{|c | c | c | c  |}
	\rowcolor{lightgray}
	\hline
	Target property  & 1 & 2 & 3  \\
	log(K$_{H}$) CO$_{2}$ & electronegativity & Poisson's ratio & Mendeleev's number \\
	log(K$_{H}$) CH$_{4}$ & electronegativity & Poisson's ratio & thermal conductivity  \\
	5.8 bar CH$_{4}$ & thermal conductivity & Poisson's ratio & Brinell's hardness  \\
	65 bar CH$_{4}$ & thermal conductivity & electronegativity & melting point  \\
	\hline
	\end{tabular}
	\caption{\textbf{Material properties sharing overlap with word embedding feature importances.} Machine learning models trained with elemental word embeddings and materials properties are compared to the models trained with MOF composition word embeddings and MOF target properties for the CoREMOF dataset. The feature importances of each model are analyzed, and compared by Jaccard similarity. The top three materials properties most similar to the model trained to MOF target properties are listed.}
	\label{tab:we-interpret}
\end{table}
We focus on
interpreting the results from a MOF design perspective. The
word-embedding features play a bigger role than topology in predicting
log(K$_{H}$) CO$_{2}$
(\cref{fig:coremofid_metrics}). For this target, the machine learning
model trained on electronegativity was the most similar to the
model trained on the word embeddings for each MOF. This suggests
that local interactions are more significant in carbon dioxide
adsorption in the infinite dilution regime, which is consistent with qualitative descriptions of
low pressure or dilute-limit profiles of absorptivity in porous materials from
literature \cite{moosavi2020}. 

Thermal conductivity also appears multiple times,
and is the most relevant elemental property for high pressure
\ce{CH4} adsorption. The relevance of thermal conductivity at higher pressures is
more difficult to interpret, given that thermal conductivity contains an
electronic and vibrational component. However, a relationship between thermal
conductivity and MOF geometry has been suggested previously.  Specifically,
thermal conductivity correlates with pore size and porosity
\cite{sumirat_theoretical_2006, wilmer2017}, which in turn affects adsorption.
Thus, when designing a MOF, including or substituting metal atoms which have low thermal conductivity in their phase pure form may improve adsorption in MOF structures. 
The coordination environment and identity of the coordinating linkers also likely plays a role in determining the trend for a given site. For reference, we have included the compositions of the high adsorption MOFs for each prediction task in the supplementary material.

Another materials
property that appeared multiple times for multiple MOF targets was the
Poisson's ratio, which reflects elasticity of a material. This is another
property that fits in the existing paradigm of MOF design: namely, flexibility.  MOFs
with flexible frameworks often are better
adsorbents \cite{doi:10.1021/acs.chemmater.5b00046}, since they can accommodate a
larger space to fit a gas molecule with less stress.

In summary, the latent information contained in the word embeddings overlaps
with known descriptors for MOF gas adsorption, pointing to important chemical
features for designing high adsorption MOFs \cite{lee2017crystallization}. 



\section{Conclusions}

We have developed an automated end-to-end machine learning framework for 
MOFs, and nanoporous materials in general, by using persistent homology and word embeddings.
Our approach builds a complex and holistic representation of the materials using only the basic input material structure,
requiring less handcrafting and domain expert guidance
than the currently widely--used porosity and chemical descriptors. 
Our topological representation is a vectorized persistence diagram,
obtained from the atomic coordinates of the normalized 
supercell representation of a materials' crystal structure. It
can be used in any machine learning algorithm. We augment the topological
information with element embeddings, constructed from a large set of scientific
abstracts via the word2vec algorithm \cite{wordembeddings2019}. They provide a
generalized representation of the MOF composition.
We have tested this approach on three different datasets, predicting
several important methane and carbon capture adsorption targets at various pressures.
These experiments show a significantly improved performance compared to standard structural
descriptors. The topological features we compute are generic and transferable across
different property targets. As the topological descriptors consistently 
outperform standard structural descriptors, they provide 
a simple way to boost the performance of any machine learning 
algorithm. Additionally, to our knowledge, these descriptors 
are the best purely geometric descriptors for predicting gas 
adsorption at low pressures and in the infinite dilution regime. Moreover,
these descriptors are interpretable: their components can be traced
to specific channels and voids in the crystal structure, which contributes
to a greater understanding of structure--property relationships in MOFs.


We conclude by highlighting the key strengths of our approach.
1) It is an ML pipeline that can automatically generate descriptors for a particular material's
prediction task without the need to handcraft specific features.
We make large gains in performance (ranging from an average 25--30\% in root-mean-square-deviation and an average 45--50\% increase in R$^2$ scores) across numerous different gas adsorption targets.
2) The generalizability and transferability of our ML model provides a way to quickly screen any dataset to find the top MOFs
for a particular task without the need to handcraft specific features, speeding up high-throughput screening of materials for adsorption
applications. As our results show, topological descriptors should be used for any porous materials adsorption prediction task and bring us closer to having a universal predictor for adsorption in porous materials.
3) Our model helps guide materials design by directly connecting property
predictions to the crystal structure, thereby encouraging the targeted design of structures to maximize desirable properties.


\section{Acknowledgements}
This work was supported by the U.S. Department of Energy under Contract Number DE-AC02-05CH11231 at Lawrence Berkeley National Laboratory. A.S.K. was supported by the Alvarez Fellowship in the Computational Research Division at LBNL. A.S.K. also acknowledges support from the TRI-AMDD internship program and further support from TRI-AMDD for cloud computing resources and initial project planning. The authors acknowledge helpful discussions with Muratahan Aykol, John Dagdelen, Matthew Horton, Aayush Singh, Ram Seshadri, and Santosh Suram.
This research used resources of the National Energy Research Scientific Computing Center (NERSC), a U.S. Department of Energy Office of Science User Facility located at Lawrence Berkeley National Laboratory, operated under Contract No. DE-AC02-05CH11231.



\section{Code availability} 

The code for generation of the material representations is available at: \url{http://www.github.com/a1k12/molecule-tda}.

\section{Competing interests}

The authors declare no competing interests.

\bibliographystyle{unsrt}
\bibliography{references}

\clearpage

\begin{center}
 {\huge Supplementary Information} 
\end{center}

\begin{center}
 \author{Aditi S. Krishnapriyan, Joseph Montoya, Maciej Haranczyk, Jens Hummelshøj, Dmitriy Morozov}
 \end{center}

\setcounter{figure}{0}  
\setcounter{table}{0}  
\setcounter{section}{0} 


\pgfplotscreateplotcyclelist{barlist}
{
    {black,fill=magenta!60!white},
    {black,fill=blue!60!white},
    {black,fill=red!60!white},
    {black,fill=orange!60!white},
    {black,fill=green!60!white},
    {black,fill=cyan!60!white},
}

\begin{figure}[H]
\begin{minipage}{0.9\textwidth}
\centering
\pgfplotstableread{feature_comparisons/rmsd_bw20k_mofs_normalized_v2.txt}{\mytable}
\begin{tikzpicture} 
\begin{axis}[
                ybar,bar width=6pt,
                symbolic x coords={logKH_CO2, logKH_CH4, pure_uptake_CO2_298.00_15000, pure_uptake_CO2_298.00_1600000, CH4LPSTP, CH4HPSTP},
                xtick={logKH_CO2, logKH_CH4, pure_uptake_CO2_298.00_15000, pure_uptake_CO2_298.00_1600000, CH4LPSTP, CH4HPSTP},
                xticklabels={log(K$_{H}$) CO$_2$, log(K$_{H}$) CH$_4$, 0.15 bar CO$_2$, 16 bar CO$_2$, 5.8 bar CH$_4$, 65 bar CH$_4$},
                xticklabel style={rotate=75, anchor=east, align=right},
                height=2.5in,
                width=.99\textwidth,
                ymin=0,
                ytick={1},
                yticklabels={Max},
                ylabel={Normalized RMSD, BW},
                legend to name=rmsd-norm-legend,
                legend style={draw=none},
                legend columns=3, transpose legend,
                legend cell align={left},
                cycle list name=barlist,
                nodes near coords,
                point meta=explicit,
                every node near coord/.append style={rotate=90, anchor=west},
                nodes near coords style={font=\footnotesize},
                ]
    \addplot    +[] table[x=target, y=geometric-topology-we-bw20k-normalized,meta=geometric-topology-we-bw20k]{feature_comparisons/rmsd_bw20k_mofs_normalized_v4.txt};
    \addlegendentry{T + S + WE}
    \addplot    +[] table[x=target, y=combo-bw20k-normalized,meta=combo-bw20k]          {feature_comparisons/rmsd_bw20k_mofs_normalized_v4.txt};
    \addlegendentry{T + WE}
    \addplot    +[] table[x=target, y=geometric-topology-bw20k-normalized,meta=geometric-topology-bw20k]{feature_comparisons/rmsd_bw20k_mofs_normalized_v4.txt};
    \addlegendentry{T + S}
    \addplot    +[] table[x=target, y=topology-bw20k-normalized,meta=topology-bw20k]    {feature_comparisons/rmsd_bw20k_mofs_normalized_v4.txt};
    \addlegendentry{Topology}
    \addplot    +[] table[x=target, y=geometric-bw20k-normalized,meta=geometric-bw20k]{feature_comparisons/rmsd_bw20k_mofs_normalized_v4.txt};
    \addlegendentry{Structural}
    \addplot    +[] table[x=target, y=we-bw20k-normalized,meta=we-bw20k]                {feature_comparisons/rmsd_bw20k_mofs_normalized_v4.txt};
    \addlegendentry{Word Embeddings}
    \path
    (axis cs:logKH_CH4, \pgfkeysvalueof{/pgfplots/ymin})
    -- coordinate (tmpmin)
    (axis cs: pure_uptake_CO2_298.00_15000, \pgfkeysvalueof{/pgfplots/ymin})
    (axis cs: logKH_CH4, \pgfkeysvalueof{/pgfplots/ymax})
    -- coordinate (tmpmax)
    (axis cs: pure_uptake_CO2_298.00_15000, \pgfkeysvalueof{/pgfplots/ymax});
  \draw[thick, dashed] (tmpmin) -- (tmpmax);
     \path
    (axis cs:pure_uptake_CO2_298.00_1600000, \pgfkeysvalueof{/pgfplots/ymin})
    -- coordinate (tmpmin)
    (axis cs: CH4LPSTP, \pgfkeysvalueof{/pgfplots/ymin})
    (axis cs: pure_uptake_CO2_298.00_1600000, \pgfkeysvalueof{/pgfplots/ymax})
    -- coordinate (tmpmax)
    (axis cs: CH4LPSTP, \pgfkeysvalueof{/pgfplots/ymax});
    \draw[thick, dashed] (tmpmin) -- (tmpmax);
\end{axis}
\end{tikzpicture}
\end{minipage}

\hfill 

\begin{minipage}{0.9\textwidth}
\centering
\begin{tikzpicture} 
\begin{axis}[   ybar,bar width=6pt,
                symbolic x coords={logKH_CO2, logKH_CH4, pure_uptake_CO2_298.00_15000, pure_uptake_CO2_298.00_1600000, CH4LPSTP, CH4HPSTP},
                xtick={logKH_CO2, logKH_CH4, pure_uptake_CO2_298.00_15000, pure_uptake_CO2_298.00_1600000, CH4LPSTP, CH4HPSTP},
                xticklabels={log(K$_{H}$) CO$_2$, log(K$_{H}$) CH$_4$, 0.15 bar CO$_2$, 16 bar CO$_2$, 5.8 bar CH$_4$, 65 bar CH$_4$},
                xticklabel style={rotate=75, anchor=east, align=right},
                height=2.5in,
                width=.99\textwidth,
                ylabel={\Rtwo score, BW},
                ymax=1,
                cycle list name=barlist,
                nodes near coords,
                every node near coord/.append style={rotate=90, anchor=west},
                nodes near coords style={font=\footnotesize}
                ]
 
    \addplot    +[] table[x=target, y=geometric-topology-we-bw20k]                {feature_comparisons/r2_bw20k_mofs_v4.txt}; 
    \addplot    +[] table[x=target, y=combo-bw20k]                   {feature_comparisons/r2_bw20k_mofs_v4.txt}; 
    \addplot    +[] table[x=target, y=geometric-topology-bw20k]                {feature_comparisons/r2_bw20k_mofs_v4.txt}; 
    \addplot    +[] table[x=target, y=topology-bw20k]                {feature_comparisons/r2_bw20k_mofs_v4.txt}; 
    \addplot    +[] table[x=target, y=geometric-bw20k]                {feature_comparisons/r2_bw20k_mofs_v4.txt}; 
    \addplot    +[] table[x=target, y=we-bw20k]                {feature_comparisons/r2_bw20k_mofs_v4.txt}; 
    \path
    (axis cs:logKH_CH4, \pgfkeysvalueof{/pgfplots/ymin})
    -- coordinate (tmpmin)
    (axis cs: pure_uptake_CO2_298.00_15000, \pgfkeysvalueof{/pgfplots/ymin})
    (axis cs: logKH_CH4, \pgfkeysvalueof{/pgfplots/ymax})
    -- coordinate (tmpmax)
    (axis cs: pure_uptake_CO2_298.00_15000, \pgfkeysvalueof{/pgfplots/ymax});
  \draw[thick, dashed] (tmpmin) -- (tmpmax);
     \path
    (axis cs:pure_uptake_CO2_298.00_1600000, \pgfkeysvalueof{/pgfplots/ymin})
    -- coordinate (tmpmin)
    (axis cs: CH4LPSTP, \pgfkeysvalueof{/pgfplots/ymin})
    (axis cs: pure_uptake_CO2_298.00_1600000, \pgfkeysvalueof{/pgfplots/ymax})
    -- coordinate (tmpmax)
    (axis cs: CH4LPSTP, \pgfkeysvalueof{/pgfplots/ymax});
    \draw[thick, dashed] (tmpmin) -- (tmpmax);
\end{axis}
\end{tikzpicture}
\end{minipage}
\begin{center}
\ref{rmsd-norm-legend}
\end{center}
\vspace*{-3mm}
\caption{\textbf{Model performances on BW dataset.} Comparison of root-mean-square deviation (left),
         coefficient of determination (right) in predicting the Henry's coefficient (log k$_{H}$) for CO$_{2}$ and CH$_{4}$, gas uptakes for CO$_{2}$, and gas uptakes for CH$_{4}$, for different features for the BW20K dataset. For each target, the units are mol kg$^{-1}$ Pa$^{-1}$, mmol/g, and VSTP/V respectively. Due to the difference in units between targets, RMSD values are normalized with respect to the maximum value in each category. The black, dashed line defines the categories that share the same units. In the case of RMSD, these categories also share a normalization factor. The actual RMSD and \Rtwo values are shown above each bar.
         }
\label{fig:bw20k_metrics}
\end{figure}
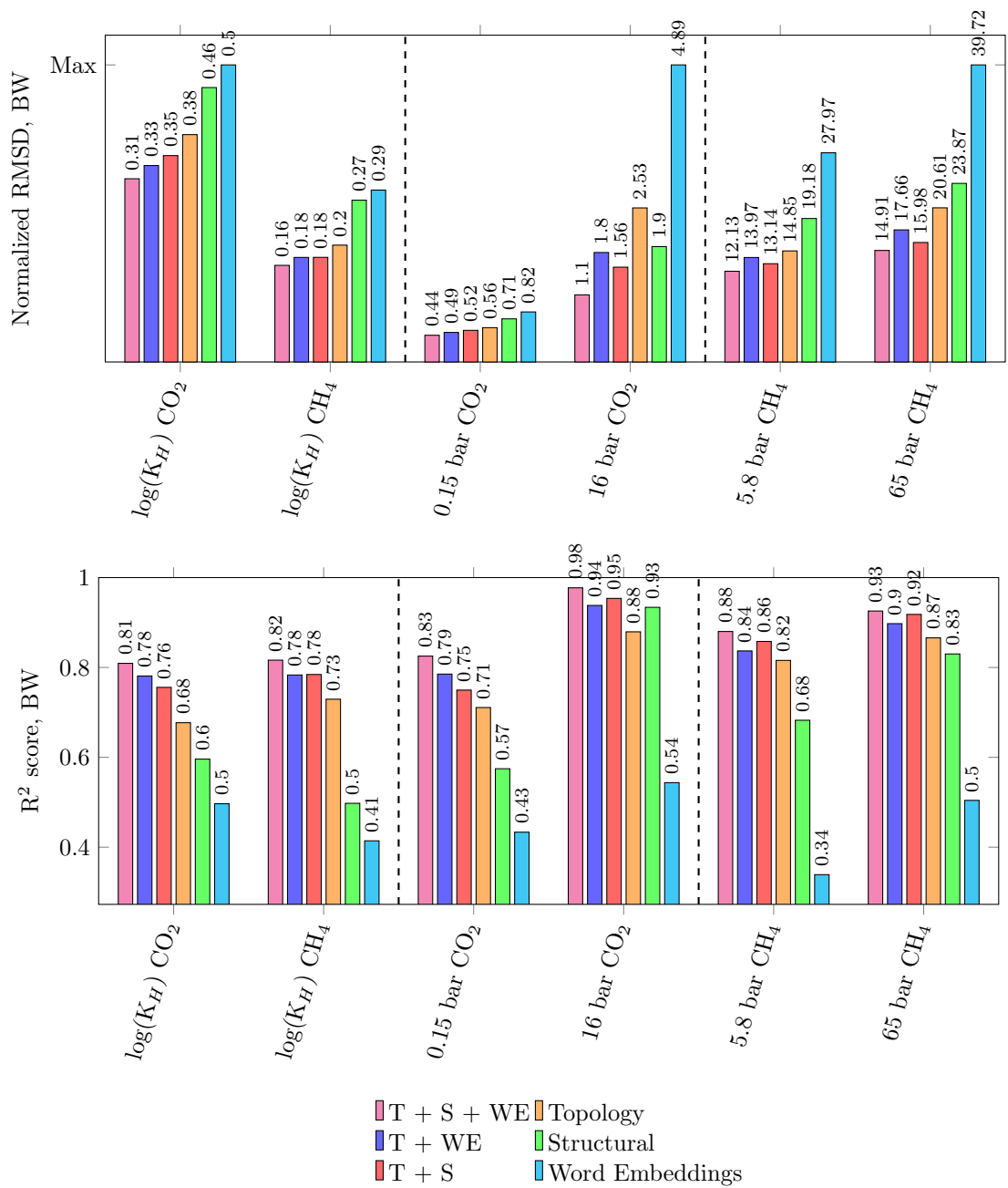


\begin{figure}[H]
\begin{minipage}{0.9\textwidth}
\centering
\begin{tikzpicture} 
\begin{axis}[
                ybar,bar width=6pt,
                symbolic x coords={logKH_CO2, logKH_CH4, CH4LPSTP, CH4HPSTP},
                xtick={logKH_CO2, logKH_CH4, CH4LPSTP, CH4HPSTP},
                xticklabels={log(K$_{H}$) CO$_2$, log(K$_{H}$) CH$_4$, 5.8 bar CH$_4$, 65 bar CH$_4$},
                xticklabel style={rotate=75, anchor=east, align=right},
                height=2.5in,
                width=.99\textwidth,
                ymin=0,
                ytick={1},
                yticklabels={Max},
                legend columns = 2,
                ylabel={Normalized RMSD, CoREMOFs},
                legend to name=rmsd-norm-legend-core,
                legend style={draw=none},
                legend columns=3, transpose legend,
                legend cell align={left},
                cycle list name=barlist,
                nodes near coords,
                point meta=explicit,
                every node near coord/.append style={rotate=90, anchor=west},
                nodes near coords style={font=\footnotesize},
                ]

    \addplot    +[] table[x=target, y=geometric-topology-we-core-normalized,meta=geometric-topology-we-core]{feature_comparisons/rmsd_core_mofid_normalized_4targets.txt};
    \addlegendentry{T + S + WE}
    \addplot    +[] table[x=target, y=combo-core-normalized,meta=combo-core]          {feature_comparisons/rmsd_core_mofid_normalized_4targets.txt};
    \addlegendentry{T + WE}
    \addplot    +[] table[x=target, y=geometric-topology-core-normalized,meta=geometric-topology-core]                {feature_comparisons/rmsd_core_mofid_normalized_4targets.txt};
    \addlegendentry{T + S}
    \addplot    +[] table[x=target, y=topology-core-normalized,meta=topology-core]    {feature_comparisons/rmsd_core_mofid_normalized_4targets.txt};
    \addlegendentry{Topology}
    \addplot    +[] table[x=target, y=geometric-core-normalized,meta=geometric-core]{feature_comparisons/rmsd_core_mofid_normalized_4targets.txt};
    \addlegendentry{Structural}
    \addplot    +[] table[x=target, y=we-core-normalized,meta=we-core]                {feature_comparisons/rmsd_core_mofid_normalized_4targets.txt};
    \addlegendentry{Word Embeddings}
    \path
    (axis cs:logKH_CH4, \pgfkeysvalueof{/pgfplots/ymin})
    -- coordinate (tmpmin)
    (axis cs: CH4LPSTP, \pgfkeysvalueof{/pgfplots/ymin})
    (axis cs: logKH_CH4, \pgfkeysvalueof{/pgfplots/ymax})
    -- coordinate (tmpmax)
    (axis cs: CH4LPSTP, \pgfkeysvalueof{/pgfplots/ymax});
  \draw[thick, dashed] (tmpmin) -- (tmpmax);
\end{axis}
\end{tikzpicture}
\end{minipage}
\begin{minipage}{0.9\textwidth}
\centering
\begin{tikzpicture} 
\begin{axis}[   ybar,bar width=6pt,
                symbolic x coords={logKH_CO2, logKH_CH4, CH4LPSTP, CH4HPSTP},
                xtick={logKH_CO2, logKH_CH4, CH4LPSTP, CH4HPSTP},
                xticklabels={log(K$_{H}$) CO$_2$, log(K$_{H}$) CH$_4$, 5.8 bar CH$_4$, 65 bar CH$_4$},
                xticklabel style={rotate=75, anchor=east, align=right},
                height=2.5in,
                width=.99\textwidth,
                ylabel={\Rtwo score, CoREMOFs},
                cycle list name=barlist,
                nodes near coords,
                every node near coord/.append style={rotate=90, anchor=west},
                nodes near coords style={font=\footnotesize},
                ]
    \addplot    +[] table[x=target, y=geometric-topology-we-core]   {feature_comparisons/r2_core_mofid_4targets.txt}; 
    \addplot    +[] table[x=target, y=combo-core]                   {feature_comparisons/r2_core_mofid_4targets.txt}; 
    \addplot    +[] table[x=target, y=geometric-topology-core]      {feature_comparisons/r2_core_mofid_4targets.txt}; 
    \addplot    +[] table[x=target, y=topology-core]                {feature_comparisons/r2_core_mofid_4targets.txt}; 
    \addplot    +[] table[x=target, y=geometric-core]               {feature_comparisons/r2_core_mofid_4targets.txt}; 
    \addplot    +[] table[x=target, y=we-core]                      {feature_comparisons/r2_core_mofid_4targets.txt}; 
    \path
    (axis cs:logKH_CH4, \pgfkeysvalueof{/pgfplots/ymin})
    -- coordinate (tmpmin)
    (axis cs: CH4LPSTP, \pgfkeysvalueof{/pgfplots/ymin})
    (axis cs: logKH_CH4, \pgfkeysvalueof{/pgfplots/ymax})
    -- coordinate (tmpmax)
    (axis cs: CH4LPSTP, \pgfkeysvalueof{/pgfplots/ymax});
  \draw[thick, dashed] (tmpmin) -- (tmpmax);
\end{axis}
\end{tikzpicture}
\end{minipage}
\begin{center}
\ref{rmsd-norm-legend-core}
\end{center}
\caption{\textbf{Model performances on CoREMOF dataset.} Comparison of root-mean-square deviation (left),
         coefficient of determination (right) in predicting the Henry's coefficient(log k$_{H}$) for CO$_{2}$ and CH$_{4}$ and gas uptakes for CH$_{4}$, for different features for the CoREMOF dataset. For each target, the units are mol kg$^{-1}$ Pa$^{-1}$ and VSTP/V respectively. Due to the difference in units between targets, RMSD values are normalized with respect to the maximum value in each category. The black, dashed line defines the categories that share the same units. In the case of RMSD, these categories also share a normalization factor. The actual RMSD and \Rtwo values are shown above each bar.
         }
\label{fig:coremofid_metrics}
\end{figure}

\begin{figure}[H]
\centering
\pgfplotsset
{
    /pgfplots/colormap={viridis-reversed-white}{%
      rgb=(1.,1.,1.)
      rgb=(0.99324,0.90616,0.14394)
      rgb=(0.84557,0.88733,0.0997)
      rgb=(0.68895,0.86545,0.18272)
      rgb=(0.53561,0.83578,0.2819)
      rgb=(0.39517,0.79747,0.36775)
      rgb=(0.27415,0.75198,0.4366)
      rgb=(0.18065,0.7014,0.48819)
      rgb=(0.12808,0.64775,0.5235)
      rgb=(0.12115,0.59274,0.54465)
      rgb=(0.13777,0.53749,0.5549)
      rgb=(0.1592,0.48224,0.55807)
      rgb=(0.18225,0.42618,0.55711)
      rgb=(0.20862,0.36775,0.55267)
      rgb=(0.23744,0.3052,0.54192)
      rgb=(0.26366,0.23763,0.51877)
      rgb=(0.28026,0.1657,0.4765)
      rgb=(0.28192,0.08966,0.41241)
      rgb=(0.267,0.00487,0.32942)
    }
}

  \begin{minipage}[t]{0.02\textwidth}
    \textbf{(a)}
  \end{minipage}
\begin{minipage}[T]{0.4\textwidth}
	\centering
	\includegraphics[width=0.9\textwidth]{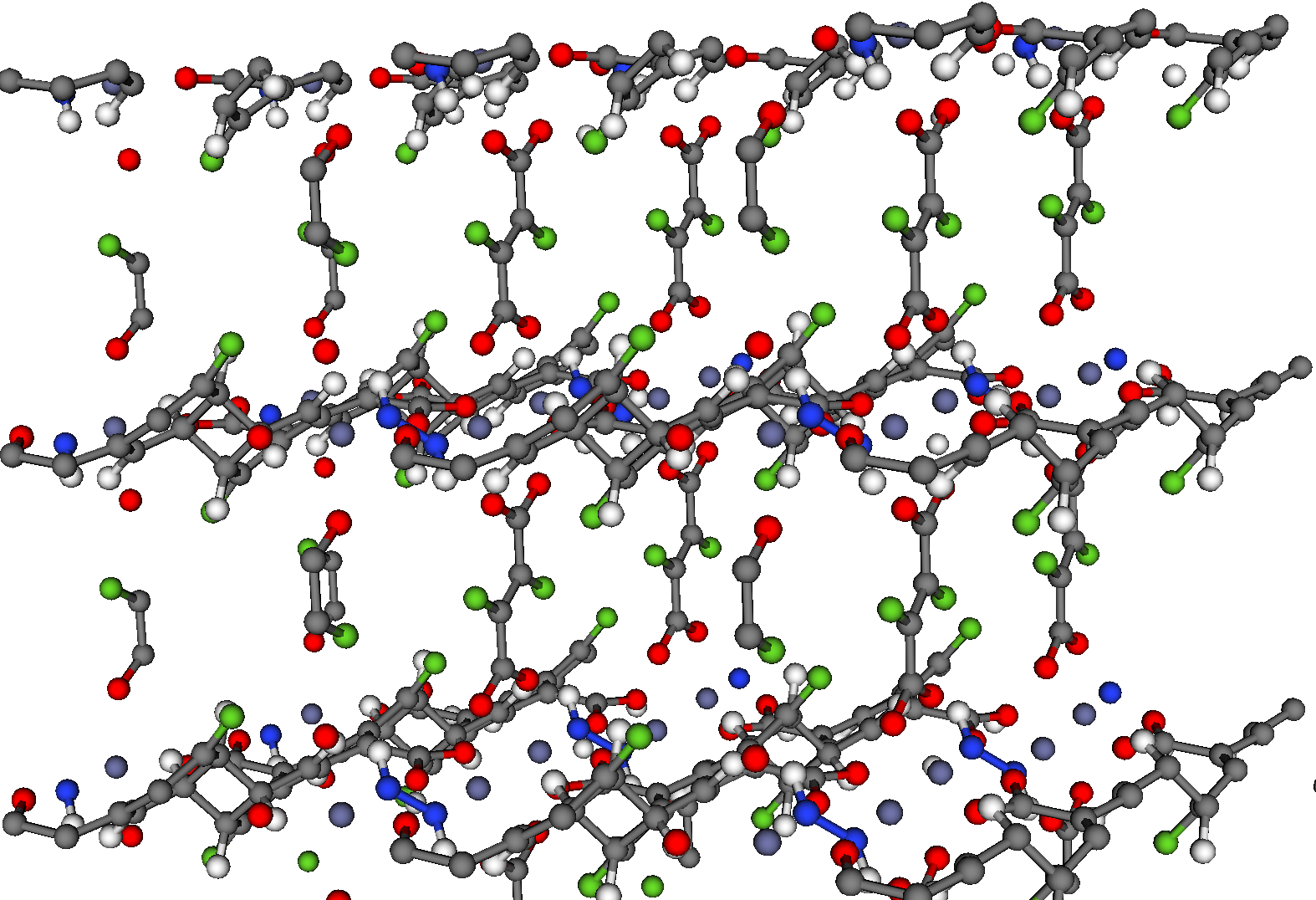}
\end{minipage}
  \begin{minipage}[t]{0.02\textwidth}
    \textbf{(b)}
  \end{minipage}
\begin{minipage}[T]{0.4\textwidth}
\begin{tikzpicture}[baseline]
\centering
\begin{axis}[
        view={0}{90},
        height=\axisdefaultheight*.75,
        xlabel=Birth (\AA),
        ylabel=Persistence (\AA),
        x coord trafo/.code={\pgfmathparse{(#1)/50*5}},
        y coord trafo/.code={\pgfmathparse{(#1)/50*2}},
    ]
    \addplot3[surf, shader=flat corner] file {feature_importances/topology_bw20k_pure_uptake_co2_298.00_15000_restrictedpi_2d.txt};
\end{axis}
\end{tikzpicture}
\end{minipage}

\hfill

  \begin{minipage}[t]{0.02\textwidth}
   \textbf{(c)}
  \end{minipage}
\begin{minipage}[T]{0.4\textwidth}
\centering
\begin{tikzpicture}
\begin{axis}
    [height=\axisdefaultheight*.75,
        		table/col sep=comma,
                xlabel=Birth (\AA),
                ylabel=Persistence (\AA),
                xmax=5,
                ymax=2]
        \addplot[
                color=blue,
                only marks,
                mark=*
                ]
         table[x=x,y=y]
            {pers_diags/str_m3_o10_o15_pcu_sym.49_pure_uptake_co2_298.00_15000_pd_2d_bvsp.txt};

    \addplot[orange, mark=*] coordinates {(3.4, 0.64)};
    \draw[orange, semitransparent, dashed] (3.4, 0.64) circle[radius=4.5pt];
    \end{axis}
\end{tikzpicture}

\end{minipage}
  \begin{minipage}[t]{0.02\textwidth}
    \textbf{(d)}
  \end{minipage}
\begin{minipage}[T]{0.4\textwidth}
	\centering
	\includegraphics[width=0.9\textwidth]{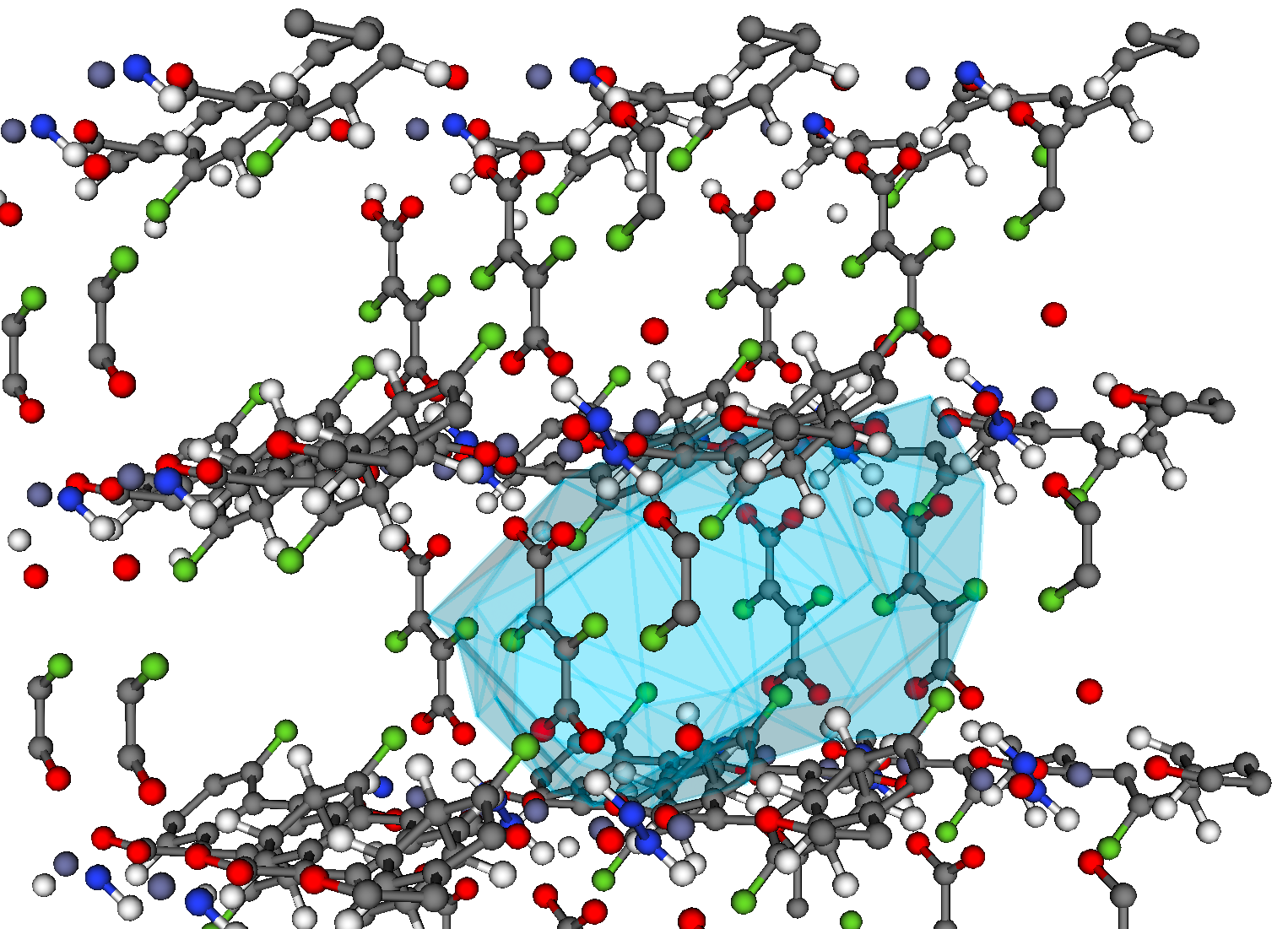} \\
\end{minipage}

\begin{minipage}[t]{\textwidth}
\centering
\begin{tikzpicture}
\pgfplotscolorbardrawstandalone[ 
    colorbar horizontal,
    point meta=explicit symbolic,
    point meta min=0,
    point meta max=1,
    xticklabel style={opacity=0,overlay},
    colorbar style={
        width=10cm,
        xtick={0,1},
    }]
    \path (10,-.55) node[right] {max} (0,-.55) node[left] {0};
\end{tikzpicture}
\end{minipage}
\caption{\textbf{Schematic outlining identification of a representative cycle in a crystal structure.} (a) The original crystal structure for MOF str-m3-o10-o15-pcu-sym.49 from the BW dataset. (b) The feature importance, shown as an image, for the 2D topology features for CO$_2$ adsorption at 0.15 bar. The color bar for this figure is shown at the bottom. (c) The 2D birth vs. persistence plot for str-m3-o10-o15-pcu-sym.49, with the (birth, persistence) point with the highest feature importance (as determined by the machine learning algorithm) in orange. The dashed lines around the orange point show the Gaussian spread factor. (d) The representative cycle for the closest point in the birth vs. persistence diagram to the orange point (specifically, the closest point that overlaps with the orange point) in (c). Figure created with VisIt 3.1.4 ({\url{https://wci.llnl.gov/simulation/computer-codes/visit}}).}
\label{fig:rep-cycle-schematic}
\end{figure}

\section{High adsorption MOFs and composition}

\subsection{hMOFs dataset}

\begin{table}[H]
\centering
\begin{tabular}{ll}
\toprule
 Structure &           Composition \\
\midrule
 hMOF-2279 &      (Zn, H, C, O, F) \\
 hMOF-3250 &   (Zn, H, C, N, O, F) \\
    hMOF-4 &      (Zn, H, C, O, F) \\
  hMOF-673 &      (Zn, H, C, O, F) \\
  hMOF-675 &      (Zn, H, C, O, F) \\
 hMOF-1722 &      (Zn, H, C, O, F) \\
 hMOF-2633 &      (Zn, H, C, O, F) \\
 hMOF-2638 &      (Zn, H, C, O, F) \\
  hMOF-469 &     (Zn, H, C, Cl, O) \\
 hMOF-2287 &      (Zn, H, C, O, F) \\
  hMOF-449 &      (Zn, H, C, O, F) \\
    hMOF-7 &      (Zn, H, C, O, F) \\
  hMOF-678 &      (Zn, H, C, O, F) \\
 hMOF-1717 &      (Zn, H, C, O, F) \\
  hMOF-953 &         (Zn, H, C, O) \\
 hMOF-3152 &  (Zn, H, C, N, Cl, O) \\
 hMOF-1721 &      (Zn, H, C, O, F) \\
  hMOF-441 &      (Zn, H, C, O, F) \\
 hMOF-1724 &      (Zn, H, C, O, F) \\
 hMOF-3142 &   (Zn, H, C, N, O, F) \\
\bottomrule
\end{tabular}
\caption{hMOFs, 0.01 bar CO$_2$}
\end{table}

\begin{table}[H]
\centering
\begin{tabular}{ll}
\toprule
 Structure &          Composition \\
\midrule
 hMOF-2279 &     (Zn, H, C, O, F) \\
 hMOF-3250 &  (Zn, H, C, N, O, F) \\
  hMOF-675 &     (Zn, H, C, O, F) \\
    hMOF-4 &     (Zn, H, C, O, F) \\
  hMOF-673 &     (Zn, H, C, O, F) \\
 hMOF-2287 &     (Zn, H, C, O, F) \\
 hMOF-1721 &     (Zn, H, C, O, F) \\
   hMOF-16 &     (Zn, H, C, O, F) \\
 hMOF-2283 &     (Zn, H, C, O, F) \\
 hMOF-1724 &     (Zn, H, C, O, F) \\
 hMOF-2633 &     (Zn, H, C, O, F) \\
  hMOF-453 &     (Zn, H, C, O, F) \\
 hMOF-2291 &     (Zn, H, C, O, F) \\
  hMOF-958 &     (Zn, H, C, O, F) \\
  hMOF-286 &        (Zn, H, C, O) \\
 hMOF-1471 &    (Zn, H, C, Cl, O) \\
 hMOF-1719 &     (Zn, H, C, O, F) \\
  hMOF-441 &     (Zn, H, C, O, F) \\
 hMOF-3246 &  (Zn, H, C, N, O, F) \\
 hMOF-2638 &     (Zn, H, C, O, F) \\
\bottomrule
\end{tabular}
\caption{0.05 bar CO$_2$}
\end{table}

\begin{table}[H]
\centering
\begin{tabular}{ll}
\toprule
 Structure &          Composition \\
\midrule
 hMOF-2279 &     (Zn, H, C, O, F) \\
 hMOF-3250 &  (Zn, H, C, N, O, F) \\
  hMOF-675 &     (Zn, H, C, O, F) \\
    hMOF-4 &     (Zn, H, C, O, F) \\
  hMOF-673 &     (Zn, H, C, O, F) \\
  hMOF-286 &        (Zn, H, C, O) \\
   hMOF-13 &     (Zn, H, C, O, F) \\
 hMOF-2291 &     (Zn, H, C, O, F) \\
 hMOF-1721 &     (Zn, H, C, O, F) \\
  hMOF-448 &     (Zn, H, C, O, F) \\
   hMOF-16 &     (Zn, H, C, O, F) \\
 hMOF-2287 &     (Zn, H, C, O, F) \\
 hMOF-3018 &        (Zn, H, C, O) \\
  hMOF-958 &     (Zn, H, C, O, F) \\
 hMOF-2283 &     (Zn, H, C, O, F) \\
 hMOF-1724 &     (Zn, H, C, O, F) \\
 hMOF-1719 &     (Zn, H, C, O, F) \\
 hMOF-2556 &     (Zn, H, C, O, F) \\
 hMOF-2342 &        (Zn, H, C, O) \\
 hMOF-3246 &  (Zn, H, C, N, O, F) \\
\bottomrule
\end{tabular}
\caption{0.1 bar CO$_2$}
\end{table}

\begin{table}[H]
\centering
\begin{tabular}{ll}
\toprule
 Structure &          Composition \\
\midrule
 hMOF-2342 &        (Zn, H, C, O) \\
 hMOF-1595 &     (Zn, H, C, O, F) \\
  hMOF-758 &     (Zn, H, C, O, F) \\
  hMOF-498 &        (Zn, H, C, O) \\
  hMOF-288 &        (Zn, H, C, O) \\
  hMOF-756 &     (Zn, H, C, O, F) \\
  hMOF-280 &        (Zn, H, C, O) \\
  hMOF-760 &     (Zn, H, C, O, F) \\
  hMOF-123 &     (Zn, H, C, O, F) \\
 hMOF-3032 &        (Zn, H, C, O) \\
 hMOF-3018 &        (Zn, H, C, O) \\
  hMOF-282 &        (Zn, H, C, O) \\
 hMOF-1507 &        (Zn, H, C, O) \\
  hMOF-754 &     (Zn, H, C, O, F) \\
 hMOF-1504 &        (Zn, H, C, O) \\
  hMOF-751 &     (Zn, H, C, O, F) \\
 hMOF-2294 &     (Zn, H, C, O, F) \\
 hMOF-2290 &     (Zn, H, C, O, F) \\
   hMOF-63 &        (Zn, H, C, O) \\
 hMOF-3257 &  (Zn, H, C, N, O, F) \\
\bottomrule
\end{tabular}
\caption{0.5 bar CO$_2$}
\end{table}

\begin{table}[H]
\centering
\begin{tabular}{ll}
\toprule
 Structure &          Composition \\
\midrule
  hMOF-237 &     (Zn, H, C, O, F) \\
 hMOF-3020 &        (Zn, H, C, O) \\
  hMOF-234 &     (Zn, H, C, O, F) \\
  hMOF-240 &     (Zn, H, C, O, F) \\
 hMOF-2293 &     (Zn, H, C, O, F) \\
  hMOF-228 &     (Zn, H, C, O, F) \\
 hMOF-2289 &     (Zn, H, C, O, F) \\
 hMOF-2277 &     (Zn, H, C, O, F) \\
 hMOF-1440 &     (Zn, H, C, O, F) \\
  hMOF-231 &     (Zn, H, C, O, F) \\
 hMOF-2832 &        (Zn, H, C, O) \\
 hMOF-2281 &     (Zn, H, C, O, F) \\
 hMOF-2285 &     (Zn, H, C, O, F) \\
 hMOF-1449 &     (Zn, H, C, O, F) \\
  hMOF-436 &     (Zn, H, C, O, F) \\
 hMOF-3248 &  (Zn, H, C, N, O, F) \\
  hMOF-439 &     (Zn, H, C, O, F) \\
 hMOF-3036 &        (Zn, H, C, O) \\
 hMOF-3244 &  (Zn, H, C, N, O, F) \\
  hMOF-834 &     (Zn, H, C, O, F) \\
\bottomrule
\end{tabular}
\caption{2.5 bar CO$_2$}
\end{table}

\subsection{BW dataset}

\begin{table}[H]
\centering
\begin{tabular}{ll}
\toprule
                  Structure &           Composition \\
\midrule
  str\_m2\_o40\_o40\_fof\_sym.19 &      (Cu, H, C, O, F) \\
  str\_m3\_o40\_o40\_fof\_sym.93 &  (Zn, H, C, Cl, O, F) \\
 str\_m5\_o10\_o29\_sra\_sym.189 &    (V, H, C, N, O, F) \\
 str\_m5\_o16\_o19\_sra\_sym.109 &       (V, H, C, O, F) \\
  str\_m3\_o40\_o40\_fof\_sym.76 &   (Zn, H, C, N, O, F) \\
  str\_m3\_o1\_o18\_pcu\_sym.101 &      (Zn, C, N, O, F) \\
  str\_m2\_o1\_o13\_pcu\_sym.157 &   (Cu, H, C, N, O, F) \\
  str\_m3\_o10\_o15\_pcu\_sym.49 &   (Zn, H, C, N, O, F) \\
   str\_m5\_o16\_o16\_sra\_sym.3 &       (V, H, C, O, F) \\
   str\_m2\_o10\_o29\_pcu\_sym.2 &   (Cu, H, C, N, O, F) \\
  str\_m3\_o40\_o40\_fof\_sym.73 &         (Zn, H, C, O) \\
   str\_m5\_o1\_o18\_sra\_sym.15 &       (V, H, C, O, F) \\
  str\_m5\_o18\_o18\_sra\_sym.27 &       (V, H, C, O, F) \\
  str\_m2\_o10\_o29\_pcu\_sym.88 &   (Cu, H, C, N, O, F) \\
 str\_m3\_o10\_o17\_pcu\_sym.151 &      (Zn, H, C, N, O) \\
  str\_m3\_o12\_o17\_pcu\_sym.61 &   (Zn, H, C, N, O, F) \\
  str\_m5\_o17\_o17\_sra\_sym.48 &      (V, H, C, Cl, O) \\
  str\_m5\_o16\_o16\_sra\_sym.11 &       (V, H, C, O, F) \\
 str\_m2\_o10\_o29\_pcu\_sym.211 &  (Cu, H, C, N, Cl, O) \\
 str\_m2\_o10\_o29\_pcu\_sym.139 &      (Cu, H, C, N, O) \\
\bottomrule
\end{tabular}
\caption{BW, 0.15 bar CO$_2$}
\end{table}

\begin{table}[H]
\centering
\begin{tabular}{ll}
\toprule
                  Structure &           Composition \\
\midrule
    str\_m7\_o3\_o3\_bcu\_sym.53 &      (Ni, H, C, N, O) \\
  str\_m3\_o11\_o29\_nbo\_sym.81 &   (Zn, H, C, N, O, F) \\
  str\_m3\_o34\_o35\_pts\_sym.67 &      (Zn, H, C, O, F) \\
  str\_m3\_o19\_o29\_nbo\_sym.12 &   (Zn, H, C, N, O, F) \\
   str\_m3\_o7\_o25\_pcu\_sym.31 &   (Zn, H, C, N, O, F) \\
    str\_m2\_o2\_o6\_pcu\_sym.58 &   (Cu, H, C, N, O, F) \\
  str\_m3\_o2\_o28\_pcu\_sym.144 &   (Zn, H, C, N, O, F) \\
  str\_m3\_o3\_o26\_pcu\_sym.176 &   (Zn, H, C, N, O, F) \\
    str\_m3\_o2\_o5\_nbo\_sym.40 &      (Zn, H, C, O, F) \\
   str\_m3\_o2\_o29\_pcu\_sym.75 &  (Zn, H, C, N, Cl, O) \\
  str\_m3\_o8\_o20\_pcu\_sym.237 &   (Zn, H, C, N, O, F) \\
  str\_m1\_o19\_o29\_pcu\_sym.37 &   (Zn, H, C, N, O, F) \\
    str\_m3\_o3\_o23\_pcu\_sym.5 &   (Zn, H, C, N, O, F) \\
 str\_m3\_o18\_o29\_pcu\_sym.124 &   (Zn, H, C, N, O, F) \\
 str\_m3\_o21\_o28\_pcu\_sym.137 &      (Zn, H, C, N, O) \\
  str\_m3\_o25\_o29\_pcu\_sym.49 &   (Zn, H, C, N, O, F) \\
  str\_m3\_o6\_o25\_pcu\_sym.247 &   (Zn, H, C, N, O, F) \\
   str\_m3\_o3\_o17\_pcu\_sym.38 &   (Zn, H, C, N, O, F) \\
  str\_m2\_o26\_o27\_pcu\_sym.21 &   (Cu, H, C, N, O, F) \\
  str\_m2\_o6\_o18\_pcu\_sym.214 &   (Cu, H, C, N, O, F) \\
\bottomrule
\end{tabular}
\caption{BW, 16 bar CO$_2$}
\end{table}

\begin{table}[H]
\centering
\begin{tabular}{ll}
\toprule
                  Structure &          Composition \\
\midrule
  str\_m5\_o18\_o18\_sra\_sym.23 &         (V, H, C, O) \\
  str\_m5\_o18\_o18\_sra\_sym.87 &      (V, H, C, N, O) \\
   str\_m5\_o18\_o18\_sra\_sym.1 &         (V, H, C, O) \\
   str\_m2\_o5\_o17\_pcu\_sym.16 &     (Cu, H, C, N, O) \\
  str\_m5\_o18\_o18\_sra\_sym.57 &         (V, H, C, O) \\
  str\_m5\_o18\_o18\_sra\_sym.38 &         (V, H, C, O) \\
  str\_m5\_o18\_o18\_sra\_sym.54 &         (V, H, C, O) \\
   str\_m3\_o10\_o15\_pcu\_sym.1 &     (Zn, H, C, N, O) \\
  str\_m5\_o18\_o18\_sra\_sym.11 &      (V, H, C, O, F) \\
   str\_m3\_o5\_o17\_pcu\_sym.24 &     (Zn, H, C, N, O) \\
 str\_m2\_o10\_o29\_pcu\_sym.144 &     (Cu, H, C, N, O) \\
   str\_m2\_o10\_o29\_pcu\_sym.3 &     (Cu, H, C, N, O) \\
  str\_m2\_o10\_o29\_pcu\_sym.20 &  (Cu, H, C, N, O, F) \\
 str\_m2\_o10\_o29\_pcu\_sym.146 &     (Cu, H, C, N, O) \\
  str\_m3\_o10\_o29\_pcu\_sym.87 &     (Zn, H, C, N, O) \\
 str\_m2\_o10\_o29\_pcu\_sym.107 &     (Cu, H, C, N, O) \\
 str\_m2\_o10\_o29\_pcu\_sym.143 &     (Cu, H, C, N, O) \\
 str\_m2\_o10\_o29\_pcu\_sym.221 &     (Cu, H, C, N, O) \\
 str\_m2\_o10\_o29\_pcu\_sym.138 &     (Cu, H, C, N, O) \\
   str\_m2\_o10\_o29\_pcu\_sym.1 &     (Cu, H, C, N, O) \\
\bottomrule
\end{tabular}
\caption{BW, 5.8 bar CH$_4$}
\end{table}

\begin{table}[H]
\centering
\begin{tabular}{ll}
\toprule
                  Structure &        Composition \\
\midrule
 str\_m3\_o40\_o41\_fof\_sym.127 &      (Zn, H, C, O) \\
  str\_m2\_o41\_o41\_fof\_sym.42 &      (Cu, H, C, O) \\
  str\_m2\_o41\_o41\_fof\_sym.24 &      (Cu, H, C, O) \\
  str\_m3\_o40\_o40\_fof\_sym.22 &      (Zn, H, C, O) \\
  str\_m2\_o40\_o41\_fof\_sym.52 &      (Cu, H, C, O) \\
  str\_m3\_o40\_o41\_fof\_sym.14 &   (Zn, H, C, N, O) \\
  str\_m2\_o41\_o41\_fof\_sym.22 &      (Cu, H, C, O) \\
   str\_m3\_o41\_o41\_fof\_sym.1 &      (Zn, H, C, O) \\
  str\_m3\_o40\_o41\_fof\_sym.27 &   (Zn, H, C, N, O) \\
  str\_m3\_o40\_o41\_fof\_sym.52 &   (Zn, H, C, N, O) \\
   str\_m3\_o40\_o41\_fof\_sym.3 &   (Zn, H, C, O, F) \\
  str\_m3\_o41\_o41\_fof\_sym.62 &      (Zn, H, C, O) \\
  str\_m3\_o40\_o41\_fof\_sym.45 &   (Zn, H, C, O, F) \\
 str\_m3\_o40\_o41\_fof\_sym.228 &      (Zn, H, C, O) \\
  str\_m3\_o17\_o17\_pcu\_sym.99 &   (Zn, H, C, N, O) \\
  str\_m3\_o40\_o41\_fof\_sym.10 &      (Zn, H, C, O) \\
  str\_m5\_o18\_o19\_sra\_sym.27 &       (V, H, C, O) \\
   str\_m2\_o40\_o41\_fof\_sym.7 &  (Cu, H, C, Cl, O) \\
  str\_m3\_o41\_o41\_fof\_sym.82 &      (Zn, H, C, O) \\
  str\_m2\_o41\_o41\_fof\_sym.46 &   (Cu, H, C, N, O) \\
\bottomrule
\end{tabular}
\caption{BW, 65 bar CH$_4$}
\end{table}

\begin{table}[H]
\centering
\begin{tabular}{ll}
\toprule
                  Structure &            Composition \\
\midrule
     str\_m4\_o4\_o5\_acs\_sym.8 &       (Cr, H, C, O, F) \\
  str\_m4\_o14\_o14\_acs\_sym.24 &       (Cr, H, C, O, F) \\
  str\_m4\_o1\_o22\_acs\_sym.197 &   (Cr, H, C, N, Cl, O) \\
   str\_m4\_o1\_o22\_acs\_sym.94 &      (Cr, C, Cl, O, F) \\
    str\_m4\_o1\_o1\_acs\_sym.10 &       (Cr, H, C, N, O) \\
 str\_m4\_o14\_o14\_acs\_sym.119 &       (Cr, H, C, O, F) \\
  str\_m4\_o4\_o15\_acs\_sym.125 &       (Cr, H, C, O, F) \\
   str\_m4\_o1\_o15\_acs\_sym.56 &    (Cr, H, C, N, O, F) \\
    str\_m4\_o1\_o1\_acs\_sym.46 &      (Cr, H, C, Cl, O) \\
  str\_m4\_o11\_o14\_acs\_sym.79 &    (Cr, H, C, S, O, F) \\
  str\_m4\_o1\_o24\_acs\_sym.165 &  (Cr, H, C, Br, Cl, O) \\
   str\_m4\_o1\_o14\_acs\_sym.68 &       (Cr, H, C, N, O) \\
   str\_m4\_o1\_o24\_acs\_sym.96 &   (Cr, H, C, N, Cl, O) \\
  str\_m4\_o12\_o15\_acs\_sym.49 &       (Cr, H, C, N, O) \\
   str\_m4\_o1\_o24\_acs\_sym.25 &      (Cr, C, Cl, O, F) \\
   str\_m4\_o4\_o15\_acs\_sym.65 &       (Cr, H, C, O, F) \\
  str\_m4\_o1\_o14\_acs\_sym.145 &   (Cr, H, C, Br, N, O) \\
   str\_m5\_o1\_o18\_sra\_sym.35 &        (V, H, C, O, F) \\
 str\_m4\_o12\_o15\_acs\_sym.129 &   (Cr, H, C, Cl, O, F) \\
  str\_m4\_o1\_o22\_acs\_sym.179 &   (Cr, H, C, Cl, O, F) \\
\bottomrule
\end{tabular}
\caption{BW, log(k$_H$) CO$_2$}
\end{table}

\begin{table}[H]
\centering
\begin{tabular}{ll}
\toprule
                  Structure &        Composition \\
\midrule
 str\_m4\_o17\_o17\_acs\_sym.152 &      (Cr, H, C, O) \\
  str\_m4\_o17\_o17\_acs\_sym.14 &      (Cr, H, C, O) \\
  str\_m4\_o17\_o17\_acs\_sym.13 &      (Cr, H, C, O) \\
 str\_m4\_o17\_o17\_acs\_sym.133 &  (Cr, H, C, Cl, O) \\
  str\_m4\_o17\_o17\_acs\_sym.99 &      (Cr, H, C, O) \\
   str\_m5\_o5\_o13\_sra\_sym.60 &       (V, H, C, O) \\
  str\_m4\_o17\_o17\_acs\_sym.82 &      (Cr, H, C, O) \\
  str\_m4\_o17\_o17\_acs\_sym.69 &      (Cr, H, C, O) \\
 str\_m4\_o17\_o17\_acs\_sym.122 &   (Cr, H, C, N, O) \\
   str\_m5\_o13\_o18\_sra\_sym.4 &       (V, H, C, O) \\
 str\_m4\_o17\_o17\_acs\_sym.171 &      (Cr, H, C, O) \\
 str\_m4\_o17\_o17\_acs\_sym.108 &      (Cr, H, C, O) \\
  str\_m4\_o17\_o17\_acs\_sym.56 &      (Cr, H, C, O) \\
  str\_m3\_o10\_o15\_pcu\_sym.23 &   (Zn, H, C, N, O) \\
  str\_m4\_o17\_o17\_acs\_sym.16 &      (Cr, H, C, O) \\
    str\_m5\_o5\_o2\_sra\_sym.49 &       (V, H, C, O) \\
 str\_m4\_o17\_o17\_acs\_sym.150 &   (Cr, H, C, S, O) \\
 str\_m4\_o17\_o17\_acs\_sym.114 &      (Cr, H, C, O) \\
 str\_m4\_o17\_o17\_acs\_sym.123 &      (Cr, H, C, O) \\
 str\_m2\_o10\_o17\_pcu\_sym.220 &   (Cu, H, C, N, O) \\
\bottomrule
\end{tabular}
\caption{BW, log(k$_H$) CH$_4$}
\end{table}

\subsection{CoREMOF dataset}

\begin{table}[H]
\centering
\begin{tabular}{ll}
\toprule
                           Structure &       Composition \\
\midrule
                        KUXSAZ\_clean &  (Fe, H, C, N, O) \\
                        TOXMUQ\_clean &     (Al, H, C, O) \\
 acs.cgd.5b01632\_ZAHKOL1438120\_clean &  (Cu, H, C, N, O) \\
                        TOXNIF\_clean &     (Al, H, C, O) \\
                        IMUVES\_clean &        (In, C, O) \\
                        XUPSAE\_clean &     (Al, H, C, O) \\
                        TOXNEB\_clean &     (Al, H, C, O) \\
                        FEVNOL\_clean &     (Cu, C, I, N) \\
                        HITXUE\_clean &        (Al, P, O) \\
                        COQNIF\_clean &    (Mn, Al, P, O) \\
                        SAPJEZ\_clean &        (Al, P, O) \\
                        YIGHIG\_clean &     (Cu, H, C, N) \\
                      PANRUS01\_clean &        (Al, P, O) \\
                        MUTGUD\_clean &     (Ag, H, C, N) \\
                      VURNED04\_clean &  (Zn, H, C, N, O) \\
                        YEMTIV\_clean &     (Cu, H, C, N) \\
                        TAKTIL\_clean &     (Ga, H, C, O) \\
                        BICPOT\_clean &     (Al, P, O, F) \\
                        KAVXAI\_clean &  (Co, H, C, N, O) \\
                        VURNED\_clean &  (Zn, H, C, N, O) \\
\bottomrule
\end{tabular}
\caption{CoREMOF, 5.8 bar CH$_4$}
\end{table}

\begin{table}[H]
\centering
\begin{tabular}{ll}
\toprule
                                    Structure &          Composition \\
\midrule
                                 YEMTER\_clean &        (Cu, H, C, N) \\
                               VURNED04\_clean &     (Zn, H, C, N, O) \\
                                 VURNED\_clean &     (Zn, H, C, N, O) \\
                                 WAFKIY\_clean &     (Zn, H, C, N, O) \\
                                 QABKIQ\_clean &     (Li, P, C, N, F) \\
                                 MOYZIK\_clean &  (V, Zn, H, C, N, O) \\
                               VURNED02\_clean &     (Zn, H, C, N, O) \\
                               VURNED01\_clean &     (Zn, H, C, N, O) \\
                                 MOKVOZ\_clean &        (Cu, H, C, O) \\
                 c6ce00465b\_c6ce00465b2\_clean &        (In, H, C, O) \\
                                 LEQCAN\_clean &        (Cu, H, C, O) \\
                                 IMUVES\_clean &           (In, C, O) \\
 acs.inorgchem.6b00661\_ic6b00661\_si\_002\_clean &           (Al, C, O) \\
                                 LAQNUP\_clean &        (Cu, H, C, O) \\
                                 KAVXAI\_clean &     (Co, H, C, N, O) \\
                                 IGORUS\_clean &     (Zn, H, C, N, O) \\
                                 CESFIQ\_clean &        (Cu, H, C, O) \\
          acs.cgd.5b01632\_ZAHKOL1438120\_clean &     (Cu, H, C, N, O) \\
                                 MUKQUG\_clean &        (Cu, H, C, O) \\
                                 LAQNOJ\_clean &        (Cu, H, C, O) \\
\bottomrule
\end{tabular}
\caption{CoREMOF, 65 bar CH$_4$}
\end{table}

\begin{table}[H]
\centering
\begin{tabular}{ll}
\toprule
                    Structure &              Composition \\
\midrule
                 DIVNIH\_clean &            (Nd, C, S, O) \\
               XEHFUL01\_clean &                (Y, C, O) \\
                 HOZDIL\_clean &         (Al, P, H, C, O) \\
                 YEZKIZ\_clean &     (Cr, Ni, H, C, N, O) \\
                 DIPMAS\_clean &            (Ce, P, C, O) \\
                 YAWKOX\_clean &            (Pr, H, C, O) \\
                 ETUWIA\_clean &                (Y, C, O) \\
                 TAGSEB\_clean &      (V, Cu, H, C, N, O) \\
                 ELIKAM\_clean &            (Pr, H, C, O) \\
                 CIKCOQ\_clean &         (Cd, H, C, N, O) \\
                 VOKJIQ\_clean &            (Al, P, H, O) \\
                 KIBDEF\_clean &            (Eu, H, C, O) \\
 c5nj02907d\_c5nj02907d2\_clean &          (V, H, C, N, O) \\
                 WAPTAL\_clean &  (Nd, As, P, H, W, C, O) \\
                 BUWMAJ\_clean &         (Nd, H, C, N, O) \\
                 XOCWET\_clean &         (La, U, H, C, O) \\
                 FATKUI\_clean &         (Ce, H, C, N, O) \\
                 DATHAJ\_clean &         (Nd, H, C, N, O) \\
                 PUMGIP\_clean &      (V, Cd, H, C, N, O) \\
                 VIWMAR\_clean &        (Nd, Cu, H, C, O) \\
\bottomrule
\end{tabular}
\caption{log(k$_H$) CO$_2$}
\end{table}

\begin{table}
\centering
\begin{tabular}{ll}
\toprule
      Structure &       Composition \\
\midrule
 ZITWIK01\_clean &           (Cd, N) \\
   ZISXEG\_clean &  (Cu, H, C, I, N) \\
   IPEBAI\_clean &  (Cu, H, C, I, N) \\
   MISQIQ\_clean &  (Al, P, H, O, F) \\
 MISQIQ02\_clean &  (Al, P, H, O, F) \\
 MISQIQ04\_clean &  (Al, P, H, O, F) \\
   VOKJIQ\_clean &     (Al, P, H, O) \\
 MISQIQ06\_clean &  (Al, P, H, O, F) \\
 MISQIQ07\_clean &  (Al, P, H, O, F) \\
 MISQIQ01\_clean &  (Al, P, H, O, F) \\
 MISQIQ05\_clean &  (Al, P, H, O, F) \\
 MISQIQ03\_clean &  (Al, P, H, O, F) \\
   GOMPOO\_clean &     (Al, P, H, O) \\
   ECIYEU\_clean &     (Al, P, H, O) \\
   XULFOA\_clean &  (Cu, H, C, N, O) \\
   VABLUG\_clean &     (Ag, H, C, N) \\
   CAHQOS\_clean &        (Al, P, O) \\
   FIJYED\_clean &        (Al, P, O) \\
   MORZID\_clean &        (Al, P, O) \\
   QOLVET\_clean &        (Al, P, O) \\
\bottomrule
\end{tabular}
\caption{log(k$_H$) CH$_4$}
\end{table}

\end{document}